\documentclass[
    10pt,
    amsmath,
    amssymb,
    prd,
    english,
    aps,
    showkeys,
]{revtex4-2}
\usepackage{fix-cm}
\usepackage[utf8]{inputenc}
\usepackage[english]{babel}
\usepackage{graphicx}
\usepackage{latexsym,amsthm,amsfonts}
\usepackage{makeidx}   
\usepackage{tensor}
\usepackage[T1]{fontenc}
\usepackage{subfigure}
\usepackage{dcolumn}
\usepackage{bm}
\usepackage{cancel}
\usepackage{xcolor}
\usepackage{hyperref}

\usepackage{textgreek}
\usepackage{comment}

\allowdisplaybreaks[1]

\usepackage{upgreek}
\usepackage{graphicx}
\usepackage{tensor}

\begin{document}
\title{Black-bounce solutions in a k-essence theory under the effects of bumblebee gravity}

\author{Carlos F. S. Pereira}
	\email{carlosfisica32@gmail.com}
	\affiliation{Departamento de Física, Universidade Federal do Espírito Santo, Av. Fernando Ferrari, 514, Goiabeiras, 29060-900, Vit\'oria, ES, Brazil.}

\author{Marcos V. de S. Silva}
	\email{marcosvinicius@fisica.ufc.br}
	\affiliation{Departamento de Física, Programa de Pós-Graduação em Física,
Universidade Federal do Ceará, Campus Pici, 60440-900, Fortaleza, Ceará, Brazil}	

\author{H. Belich} 
    \email{humberto.belich@ufes.br}
    \affiliation{Departamento de F\'isica e Qu\'imica, Universidade Federal do Esp\'irito Santo, Av.Fernando Ferrari, 514, Goiabeiras, Vit\'oria, ES 29060-900, Brazil.}
    
\author{Denis C. Rodrigues}
    \email[]{deniscr@gmail.com}
	\affiliation{Núcleo Cosmo-ufes \& Departamento de Física, Universidade Federal do Espírito Santo, Av. Fernando Ferrari, 514, Goiabeiras, 29060-900, Vit\'oria, ES, Brazil.}

	\author{Júlio C. Fabris}
    \email[]{julio.fabris@cosmo-ufes.org}
	\affiliation{Núcleo Cosmo-ufes \& Departamento de Física, Universidade Federal do Espírito Santo, Av. Fernando Ferrari, 514, Goiabeiras, 29060-900, Vit\'oria, ES, Brazil}
    \affiliation{National Research Nuclear University MEPhI (Moscow Engineering Physics Institute), 115409, Kashirskoe shosse 31, Moscow, Russia.}
	
	\author{Manuel E. Rodrigues}
    \email[]{esialg@gmail.com}
	\affiliation{Faculdade de Ci\^encias Exatas e Tecnologia, Universidade Federal do Par\'a Campus Universit\'ario de Abaetetuba, 68440-000, Abaetetuba, Par\'a, Brazil and Faculdade de F\'isica, Programa de P\'os-Gradua\c{c}\~ao em F\'isica, Universidade Federal do Par\'a, 66075-110, Bel\'em, Par\'a, Brazil.}

\begin{abstract}
\noindent In the present study, we analyze the effects of violation of Lorentz symmetry for black-bounce solutions in a $k$-essence theory that has the form of a power law for the configuration $n=1/3$. We perform such analysis for a known model explored in previous work, $\Sigma^2=x^2+a^2$, and complement the proposal with a new black-bounce model for the area functions, $\Sigma^2_1=\sqrt{x^4+d^4}$. This model has the Schwarzschild-de Sitter asymptomatic behavior for $x\to{-\infty}$, and we investigate the scalar field, potential, and energy conditions for both models. We have shown that the violation of Lorentz symmetry can be generated through $k$-essence without the need for an additional field. These results corroborate the validation of other previously investigated wormhole solutions.
\end{abstract}
	
	
\maketitle
	
\section{Introduction}\label{sec1}

Singular black hole solutions arise naturally within the theory of General Relativity (GR) and are associated with the gravitational collapse generated as the final phase of a massive star \cite{INTRO1,INTRO2,INTRO3}. In addition to the singular solutions, around the 1970s Bardeen proposed the first solution for a regular black hole \cite{INTRO4}, a black hole without singularity. This spacetime proposed by Bardeen did not satisfy the vacuum conditions. Therefore, it was only around the year 2000 that Beato and Garcia \cite{INTRO5} obtained the content of matter referring to this spacetime, belonging to a class of nonlinear electrodynamics, and the regularization parameter gained the interpretation as being the magnetic charge of the monopole. After the regular black hole solution proposed by Bardeen, several other regular models were presented in the literature and, in a certain way, expanded the understanding of these intriguing objects \cite{INTRO6,INTRO7,INTRO8,INTRO9,INTRO10,INTRO11,INTRO12,INTRO13,INTRO14,INTRO15,INTRO16,INTRO17,INTRO18}.

Following the perspective of regular solutions, Simpson and Visser recently proposed a new class of solutions called black-bounce \cite{matt}. In a straightforward way, the spherically symmetric and static Schwarzschild spacetime is considered as the starting point, and then the regularization process $x^2\to {x^2+a^2}$ is applied. Thus, this new spacetime generated does not represent a vacuum configuration, and the parameter $a^2$ can be seen as the throat of the wormhole. Depending on how the parameters of the model are related, we can have some possible types of configuration, as follows: with $a>2m$, we have a bidirectional traversable wormhole, $a=2m$, a unidirectional wormhole, and with the throat located at the origin $x=0$ and $a<2m$ we have a regular black hole with two symmetric horizons. As this model does not satisfy the vacuum conditions, some authors obtained the matter content for this spacetime, which is a combination of nonlinear electrodynamics and a ghost scalar field \cite{INTRO19, INTRO20,Rodrigues:2023vtm}. In addition to spherically symmetric and static spacetimes, new black-bounce models have been explored, for example, in the context of black-strings \cite{INTRO21,INTRO22,INTRO23,INTRO24}, modified gravity \cite{INTRO25,INTRO25A}, rotating background \cite{INTRO26,INTRO27}, teleparallel gravity \cite{INTRO28}, and conformal gravity \cite{INTRO29}.

The idea that constants in nature can vary with the evolution of the Universe was initially proposed by Dirac in the 1930s \cite{dirac1}. He suggested that the gravitational coupling constant introduced by Einstein should not remain constant on a timescale comparable to the age of the Universe. In the 1950s, Dirac \cite{dirac2} also suggested the existence of an ether that would cause anisotropy in spacetime, and then the Lorentz Symmetry Violation (LSV). In this way, another constant of nature should vary, the speed of light.

Lorentz Symmetry is one of the symmetries that led to the development of the Standard Model (SM) of Physics of fundamental interactions \cite{salam}. Although the SM has had great success in phenomenological description, it has achieved electroweak unification and predicted the Higgs boson, but it leaves out the gravitational interaction description. For this reason, an extension of this model is desirable, and a proposal inspired by spontaneous symmetry
violation (extending the background field that realizes the Higgs mechanism to a non-scalar field) causes the LSV spontaneously \cite{sam, ens}. The evolution of this research has formulated the Standard Model Extension (SME) \cite{sme,sme1, cfj}. 

In addition to the proposal of the SME, effective theories in the gravitational sector with LSV were also explored. One possibility is made if it is considered the Lorentz violation from a vector that acquires a non-zero vacuum expectation value. These theories are known as bumblebee models \cite{EXTRA}. It is known that the local LSV is accompanied by diffeomorphism violation \cite{blu}. In this scenario, the spontaneous LSV is promoted by a potential that has a minimum, which ensures the breaking of the $U(1)$ symmetry. In this context, we intend to explore this LSV in compact objects \cite{cas}.

The $k$-essence models, initially proposed to describe inflationary cosmology and later extended to explain the accelerated expansion of the Universe, introduce nonstandard kinetic terms for scalar fields coupled to gravity \cite{INTRO30,INTRO31}. In the context of compact objects, such as black holes and wormholes, $k$-essence has been used to propose exact solutions with intriguing properties, including configurations that challenge the traditional notions of event horizons and singularities \cite{denis1, denis2}. From this perspective, we study the mechanism of LSV, considering only the $k$-essence scalar field as the main ingredient, that is, without the need for an additional field, as usually happens \cite{LVN1,LVN2}.

The present study first establishes in Section \ref{sec2} the theoretical background of the $k$-essence model, including the key relationships and equations. In Sections \ref{sec3} and \ref{sec4}, we investigate two spacetimes for the $k$-essence configuration $n=1/3$ and properties of physical quantities derived from the metric function, influenced or not by the Lorentz symmetry parameter. In Section \ref{sec5}, we analyze the energy conditions for both models. Finally, Section \ref{sec6} summarizes the main conclusions of this analysis regarding the viability of constructing regular black-bounce geometries within $k$-essence theories.

\section{General relations}\label{sec2}

The $k$-essence theories are characterized by a non-canonical kinetic term for the scalar field, represented by the Lagrangian
\begin{equation}\label{Lagran}
    \mathcal{L} = \sqrt{-g}[R-F(X,\phi)]\,,
\end{equation}
where $R$ is the Ricci scalar and $X=\eta\phi_{;\rho}\phi^{;\rho}$ denotes the kinetic term. While $k$-essence models can include a potential term and non-trivial couplings, the scalar sector is generally minimally coupled to gravity. The parameter $\eta=\pm 1$ avoids imaginary terms in the kinetic expression $X$. By choosing different forms of the function $F(X,\phi)$, $k$-essence theories can describe both phantom and standard scalar fields.

The variation of the Lagrangian \eqref{Lagran} with respect to the metric and the scalar field yields the field equations
\begin{eqnarray}\label{eq1}
G_\mu^{\nu}=-T_\mu^{\nu}\left(\phi\right)=-\eta{F_X}\phi_\mu{\phi^{\nu}} + \frac{1}{2}\delta_\mu^{\nu}F, \\\label{eq2}
\eta\nabla_\alpha\left(F_X\phi^{\alpha}\right)-\frac{1}{2}F_\phi =0,
\end{eqnarray} 
where $G_\mu^{\nu}$ is the Einstein tensor, $T_\mu^{\nu}$ is the stress-energy tensor, $F_X=\frac{\partial{F}}{\partial{ X}}$, $F_\phi=\frac{\partial{F}}{\partial\phi}$, and $\phi_\mu=\partial_\mu\phi$.

The line element representing the most general spherically symmetric and static spacetime takes the form:
\begin{eqnarray}\label{eq3}
ds^2=e^{2\gamma\left(x\right)}dt^2-e^{2\alpha\left(x\right)}dx^2-e^{2\beta\left(x\right)}d\Omega^2,
\end{eqnarray} 
where $x$ is an arbitrary radial coordinate, $d\Omega^2=d\theta^2+\sin^2\theta{d\varphi^2}$ is the area element, and $\phi=\phi\left (x\right)$. 

The non-zero components of the stress-energy tensor are
\begin{eqnarray}\label{eq4}
T_0^{0}= T_2^{2} =T_3^{3}= -\frac{F}{2}, \\\label{eq5}
T_1^{1}=-\frac{F}{2} -\eta{F_X}e^{-2\alpha}{\phi}'^2,
\end{eqnarray} with $\phi'=\frac{d\phi}{dx}$.

It is assumed that the function $X=-{\eta}e^{-2\alpha}{\phi}'^2$ is positive, which implies that $\eta=-1$. As a result, the equations of motion take the forms
\begin{eqnarray}\label{eq6}
2\left(F_X{e^{-\alpha+2\beta+\gamma}}\phi'\right)' - {e^{\alpha+2\beta+\gamma}}F_\phi=0, \\\label{eq7}
{\gamma}'' + {\gamma}'\left(2{\beta}'+ {\gamma}'-{\alpha}'\right)-\frac{e^{2\alpha}}{2}\left(F-XF_X\right)=0, \\\label{eq8}
-e^{2\alpha-2\beta} + {\beta}'' +{\beta}'\left(2{\beta}'+ {\gamma}'-{\alpha}'\right) -\frac{e^{2\alpha}}{2}\left(F-XF_X\right)=0, \\\label{eq9}
-e^{-2\beta} + e^{-2\alpha}{\beta}'\left({\beta}'+2{\gamma}'\right) -\frac{F}{2} + XF_X=0.
\end{eqnarray}

Now, let us consider the line element initially derived in Refs. \cite{LV1,LV2,LV21,LV22,LV23} and applied to scattering processes in \cite{LV3}. Its general shape is defined by
\begin{eqnarray}\label{EQLV1}
    ds^2_{(g)}= \frac{\left(1-\frac{2m}{x}\right)dt^2}{\sqrt{\left(1+\frac{3W}{4}\right)\left(1-\frac{W}{4}\right)}} -\frac{dx^2}{\left(1-\frac{2m}{x}\right)}\sqrt{\frac{1+\frac{3W}{4}}{\left(1-\frac{W}{4}\right)^3}} -x^2d\Omega^2,
\end{eqnarray} where the parameter $W=\xi{b^2}$ represents the coefficient associated with the LSV. Note that in the asymptotic limit of $g_{tt}\left(x\to\infty\right)=\frac{1}{\sqrt{\left(1+\frac{3W}{4}\right)\left (1-\frac{W}{4}\right)}}$, the LSV factor modifies the asymptotic behavior of the Schwarzschild black hole. Through a redefinition of the temporal coordinate, we can rewrite the metric Eq. (\ref{EQLV1}) more pleasantly. Thus, doing ${t}\left[\left(1+\frac{3W}{4}\right)\left(1-\frac{W}{4}\right)\right]^{-1/ 4}\to{\bar{t}}$ and considering the following simplification $\Pi= \sqrt{\frac{1+\frac{3W}{4}}{\left(1-\frac{W}{ 4}\right)^3}}$, we obtain
\begin{eqnarray}\label{EQLV2}
d\bar{s}^2= \left(1-\frac{2m}{x}\right)d\bar{t}^2 -\left(1-\frac{2m}{x}\right)^{-1}\Pi{dx^2} -x^2d\Omega^2.
\end{eqnarray}

Now, using a procedure similar to Simpson-Visser \cite{matt}, we can apply the regularization process $x^2\to{x^2+a^2}$ to the metric above and then rewrite it as
\begin{eqnarray}\label{EQLV3}
    d\bar{s}^2= A(x)d\bar{t}^2- \frac{\Pi{dx^2}}{A(x)}-\Sigma^2(x)d\Omega^2,
\end{eqnarray} where $A(x)=\left(1-\frac{2m}{\sqrt{x^2+a^2}}\right)$ and $\Sigma^2(x)=x^2+a^2$, where the parameter $a$ represents the non-zero throat of the wormhole. Evaluating the functions that appear by multiplying the radial and temporal coordinates of the metric Eq. (\ref{EQLV3}), we can verify that the conditions of the horizon position are the same as those of Simpson-Visser.
An interesting aspect to note in this new metric Eq. (\ref{EQLV3}) is that the position of the event horizon $g_{xx}=0$ does not change when compared to the original work by Simpson-Visser \cite{matt}. The Simpson-Visser metric is regular in all spacetime, and in the limit where the throat radius tends to zero, $a\to{0}$, the Schwarzschild metric is recovered, however, for the metric Eq. (\ref{EQLV1}), the Schwarzschild spacetime is being modified by the LSV parameter.

The notation used here follows the same as used in the reference \cite{denis1}. We can see that the \textit{quasi-global} gauge is now being modified by the violation parameter $\alpha\left(x\right)+\gamma\left(x\right)=ln\left(\Pi\right)$. Thus, comparing the coefficients of the metrics Eq. (\ref{eq3}) and Eq. (\ref{EQLV3}), we have that $A(x) = e^{2\gamma}$, $e^{2\alpha}=\frac{\Pi}{A(x)}$, and $e^\beta = \Sigma(x)$.

In this way, making use of the new line element Eq. (\ref{EQLV3}), we can rewrite the equations of motion Eq. (\ref{eq6}-\ref{eq9}) in the new coordinates. Thus, combining Eqs. (\ref{eq7}-\ref{eq9}), we obtain the following expressions:

\begin{eqnarray}\label{eq10}
2A\frac{{\Sigma}''}{\Sigma} + XF_X\left(\Pi-2\right) + \frac{\left(1-\Pi\right)\left(2 + F\Sigma^2\right)}{\Sigma^2} + \frac{2\left(\Pi-1\right){\Sigma}'\left(A\Sigma\right)'}{\Pi\Sigma^2}=0, \\\label{eq11}
{A}''\Sigma^2 - A\left(\Sigma^2\right)''+ 2\Pi =0,
\end{eqnarray} 
where the primes now represent derivatives with respect to $x$.

The two remaining equations, Eq. (\ref{eq6}) and Eq. (\ref{eq9}), are rewritten in the new coordinates as
\begin{eqnarray}\label{eq12}
2\left(F_X{A\Sigma^2}\phi'\right)' - \Pi\Sigma^2F_\phi = 0, \\\label{eq13}
\frac{1}{\Sigma^2}\left(-\Pi + A'\Sigma'\Sigma + A{\Sigma'}^2\right) -\frac{\Pi{F}}{2} + \Pi{XF_X} = 0.
\end{eqnarray}

For the remainder of this analysis, the $k$-essence function is defined as $F(X)=F_0 X^n - 2V(\phi)$, where $F_0$ is a constant parameter, $n$ is a real number, $X=\eta \phi_{;\rho}\phi^{;\rho}$ is the kinetic term, and $V(\phi)$ represents the potential.

\section{General solution for the first model} \label{sec3}

In this section, we will use a Simpson-Visser area function $\Sigma^2\left(x\right)=x^2 +a^2$ \cite{matt,LVS}, and using the equations of motion of the $k$-essence model, we want to find out the function's corresponding metric. However, we are now considering the LSV parameter. Thus, the general solution of the differential equation (\ref{eq11}) is given by
\begin{eqnarray}\label{eq14}
    A\left(x\right)= \Pi + C_{1}\Sigma^2 + \frac{C_2\left[xa +\Sigma^2\arctan\left(\frac{x}{a}\right)\right]}{2a^3},
\end{eqnarray} where $C_1$ and $C_2$ are integration constants. It is worth noting that in the limit where $\Pi=1$, we recover the same results as in Ref. \cite{CDJM1}. As investigated in Refs. \cite{CDJM1,CDJM2}, we require that in the asymptotic limit, as $x \to \infty$, the above metric approaches the LSV parameter $\Pi$. When $\Pi$ is absent and no longer affects spacetime, it reduces to the Schwarzschild metric. Therefore, the first relationship between the integration constants is $C_1=- \frac{\pi{C_2}}{4a^3}$. The second condition occurs in the limit where $x\to{0}$. If we want our spacetime to behaves like the Simpson-Visser solution in this limit, we need that $C_2=4\left(2m + a(\Pi -1)\right)/\pi$. Thus, the general solution is defined by
\begin{eqnarray}\label{eq15}
    A\left(x\right)= \Pi + \frac{2\left(2m + a(\Pi -1)\right)}{a^3\pi}\left[xa + \Sigma^2\left(\arctan\left(\frac{x}{a}\right)-\frac{\pi}{2}\right)\right].
\end{eqnarray} It is worth noting that the position of the event horizon is modified by the LSV factor and in the limit $A(x\to{0})=1 - \frac{2m }{a}$. Therefore, when the violation parameter ceases to act, $\Pi=1$, then the position of the horizon returns to the same as Simpson-Visser when $a=2m$. We also emphasize that when performing the expansion in $x\to\infty$ of the metric function above, it behaves like a flat spacetime. However, it is not Lorentzian, depending exclusively on the violation parameter $\Pi$, which, when it assumes the value $\Pi=1$, the geometry becomes asymptotically flat and Lorentzian.

This function, as in previous works \cite{CDJM1,CDJM2}, is asymptotically flat in the limit of $x\to\infty$ when the LSV factor ceases to act and behaves similarly to the Schwarzschild-de Sitter function in the limit where $x\to{-\infty}$. This behavior can be seen in the expression below,
\begin{eqnarray}\label{eq15a}
A(x) \approx \Pi - \frac{4}{3\pi}\left[\frac{2m+a(\Pi-1)}{x}\right]-\frac{2(x^2+a^2)\left[2m+a(\Pi-1)\right]}{a^3},
\end{eqnarray} and for the asymptotic $x\to{-\infty}$ to be obeyed, it is necessary that there be the restriction $2m + a(\Pi -1)>0$.

In order to construct and analyze the Kretschmann scalar, consider the non-zero components of the Riemann tensor below, taking as a basis the spherically symmetric line element Eq. (\ref{EQLV3}),
\begin{eqnarray}\label{eq16}
\tensor{R}{^{tr}_{tr}}= \frac{A''}{2\Pi}, \quad \tensor{R}{^{\theta\phi}_{\theta\phi}}= \frac{A\Sigma'^2-\Pi}{\Pi\Sigma^2}, \quad
\tensor{R}{^{t\theta}_{t\theta}}=\tensor{R}{^{t\phi}_{t\phi}}= \frac{A'\Sigma'}{2\Pi\Sigma},  \quad  \tensor{R}{^{r\theta}_{r\theta}} =\tensor{R}{^{r\phi}_{r\phi}} = \frac{A'\Sigma' + 2A\Sigma''}{2\Pi\Sigma}. \nonumber \\
\end{eqnarray}

With the non-zero components of the Riemann tensor calculated above Eq. (\ref{eq16}), we are able to quantify the Kretschmann scalar $K=\tensor{R}{_{\alpha\beta\mu\nu}}\tensor{R}{^{\alpha\beta\mu\nu}}$, which can be defined in terms of the components of the Riemann tensor. In this way, we can write it as a positive sum of quadratic terms \cite{lobo,Bronnikov2},
\begin{eqnarray}\label{eq17}
K= 4\left( \tensor{R}{^{tr}_{tr}} \right)^2 + 4\left(\tensor{R}{^{t\theta}_{t\theta}}\right)^2 + 4\left(\tensor{R}{^{t\phi}_{t\phi}}\right)^2+ 4\left(\tensor{R}{^{r\theta}_{r\theta}}\right)^2 + 4\left(\tensor{R}{^{r\phi}_{r\phi}}\right)^2 + 4\left(\tensor{R}{^{\theta\phi}_{\theta\phi}}\right)^2.
\end{eqnarray} 

Imposing the spherical symmetry conditions, it can be written in a reduced form by the expression below,
\begin{eqnarray}\label{eq18}
K= 4\left( \tensor{R}{^{tr}_{tr}} \right)^2 + 8\left(\tensor{R}{^{t\theta}_{t\theta}}\right)^2 + 8\left(\tensor{R}{^{r\theta}_{r\theta}}\right)^2  + 4\left(\tensor{R}{^{\theta\phi}_{\theta\phi}}\right)^2.
\end{eqnarray}

Therefore, the Kretschmann scalar is defined by
\begin{eqnarray}\label{eq19}
K=\frac{({A}''\Sigma^2)^2 + 2(\Sigma{A}'{\Sigma}')^2 + 4(A{\Sigma}'^2-\Pi)^2 +2\Sigma^2({A}'{\Sigma}'+2A{\Sigma}'')^2}{\Pi^2\Sigma^4}.
\end{eqnarray}

Using the combination of Eqs. (\ref{eq10}) and (\ref{eq13}) for the $k$-essence field and the metric functions Eq. (\ref{eq15}) and $\Sigma^2(x)$, we obtain the expression $2A{\Sigma}''/\Sigma=\Pi{XF_X}$. Using the configuration of the $k$-essence field $n=1/3$, we obtain the expression for the scalar field that is defined below:
\begin{eqnarray}\label{eq20}
\phi(x)&=& \frac{C_2 D_1 x \left(a x-\pi  \left(a^2+x^2\right)\right)}{8 a^5 \left(a^2+x^2\right)^2} +\frac{D_1 \Pi  x \left(5 a^2+3 x^2\right)}{8 a^4 \left(a^2+x^2\right)^2} +\frac{C_2 D_1 \left(2 a x-\pi  \left(a^2+x^2\right)\right) \tan ^{-1}\left(\frac{x}{a}\right)}{8 a^6 \left(a^2+x^2\right)} \nonumber \\
&+&\frac{3 D_1 \Pi  \tan ^{-1}\left(\frac{x}{a}\right)}{8 a^5} + \frac{C_2 D_1 \left(\tan ^{-1}\left(\frac{x}{a}\right)\right)^2}{8 a^6},
\end{eqnarray} where the constants are defined as $D_1=(\frac{6a^2}{F_0{\Pi}^{2/3}})^{3/2}$ and $C_2=4\left(2m + a(\Pi -1)\right)/\pi$.

The asymptotic behavior when $x\to\pm\infty$ for the scalar field described above is given by

\begin{eqnarray}\label{eq21}
\phi\left(x\to\infty\right) =\frac{D_1\pi}{32a^6}\left(6a\Pi -\pi{C_2}\right), \qquad \qquad \phi\left(x\to{-\infty}\right) =\frac{D_1\pi}{32a^6}\left(3\pi{C_2}-6a\Pi\right),
\end{eqnarray} where we set $F_0=1$.

Similarly, using the equation of motion (\ref{eq13}) for the metric function defined in Eq. (\ref{eq15}) we obtain the potential function for the $k$-essence configuration defined above,
\begin{eqnarray}\label{eq21a}
V\left(\phi\left(x\right)\right)= \frac{1}{\Sigma^2} -\frac{{{\Sigma}'A'}}{\Pi\Sigma}-\frac{A{\Sigma'}^2}{\Pi\Sigma^2} + \frac{A{\Sigma}''}{\Pi\Sigma},
\end{eqnarray} where its explicit form is given below as
\begin{eqnarray}\label{eq22}
V\left(\phi(x)\right)= \frac{2 a^2}{\left(a^2+x^2\right)^2} + \frac{C_2 \left(-2 a x \left(a^2+3 x^2\right)+\pi  a^2 \left(2 x^2-a^2\right)+3 \pi  x^4\right)}{4 a^3 \Pi  \left(a^2+x^2\right)^2} +\frac{C_2 \left(a^2-3 x^2\right) \tan ^{-1}\left(\frac{x}{a}\right)}{2 a^3 \Pi  \left(a^2+x^2\right)}.
\end{eqnarray} 

To better represent the potential in Eq. (\ref{eq22}), we define the following transformation of variables $\psi=\arctan\left(\frac{x}{a}\right)$. Thus, the above potential is rewritten in terms of this new variable $\psi$ as
\begin{eqnarray}\label{eq23}
    V\left(\psi\right)= \left(\frac{3 \pi  C_2}{4 a^3 \Pi } \right) + \frac{2 \cos ^4(\psi )}{a^2} -\left(\frac{3 C_2}{2 a^3 \Pi } \right)\psi +\frac{C_2 (2 \psi -\pi )}{a^3 \Pi }\cos^2\left(\psi\right) +\left(\frac{C_2}{4 a^3 \Pi }\right)\sin\left(2\psi\right)\left(\cos\left(2\psi\right)-2\right).
\end{eqnarray}

\begin{figure}[htb!]
\centering  
	\mbox{
	\subfigure[]{\label{A1MOD1}
	{\includegraphics[width=0.45\linewidth]{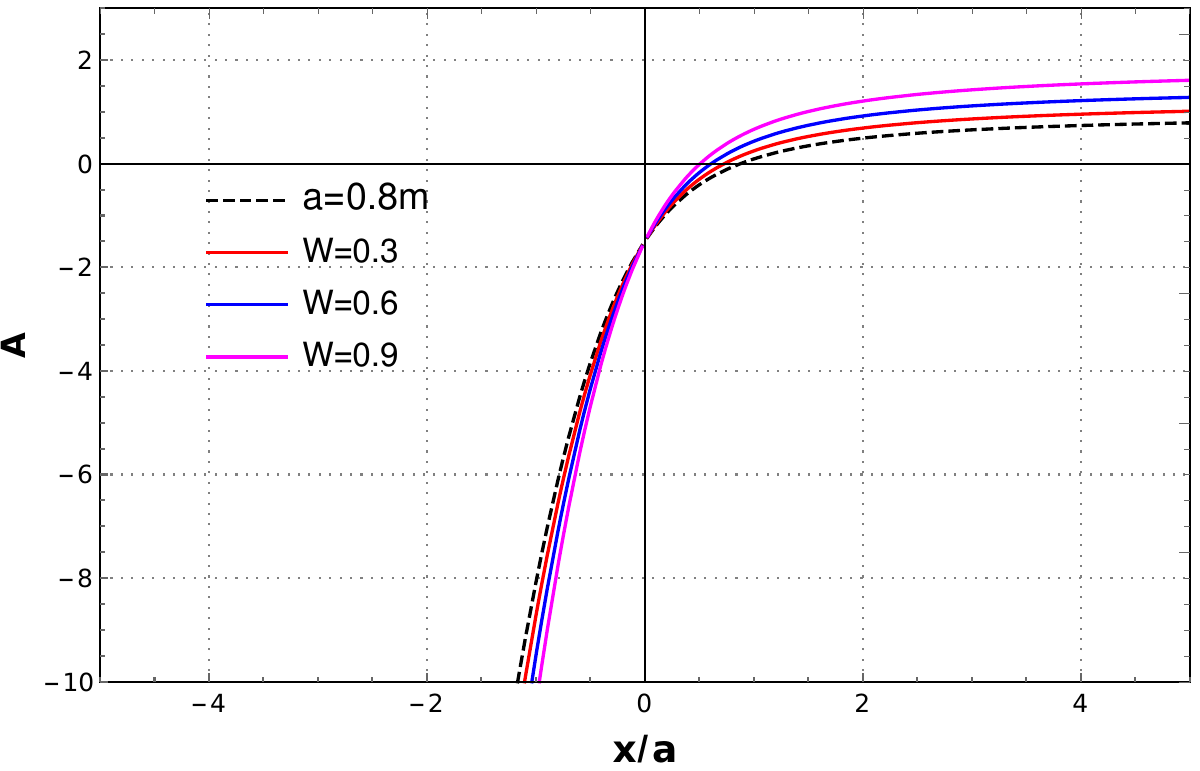}}}\qquad
	\subfigure[]{\label{A2MOD1}
	{\includegraphics[width=0.45\linewidth]{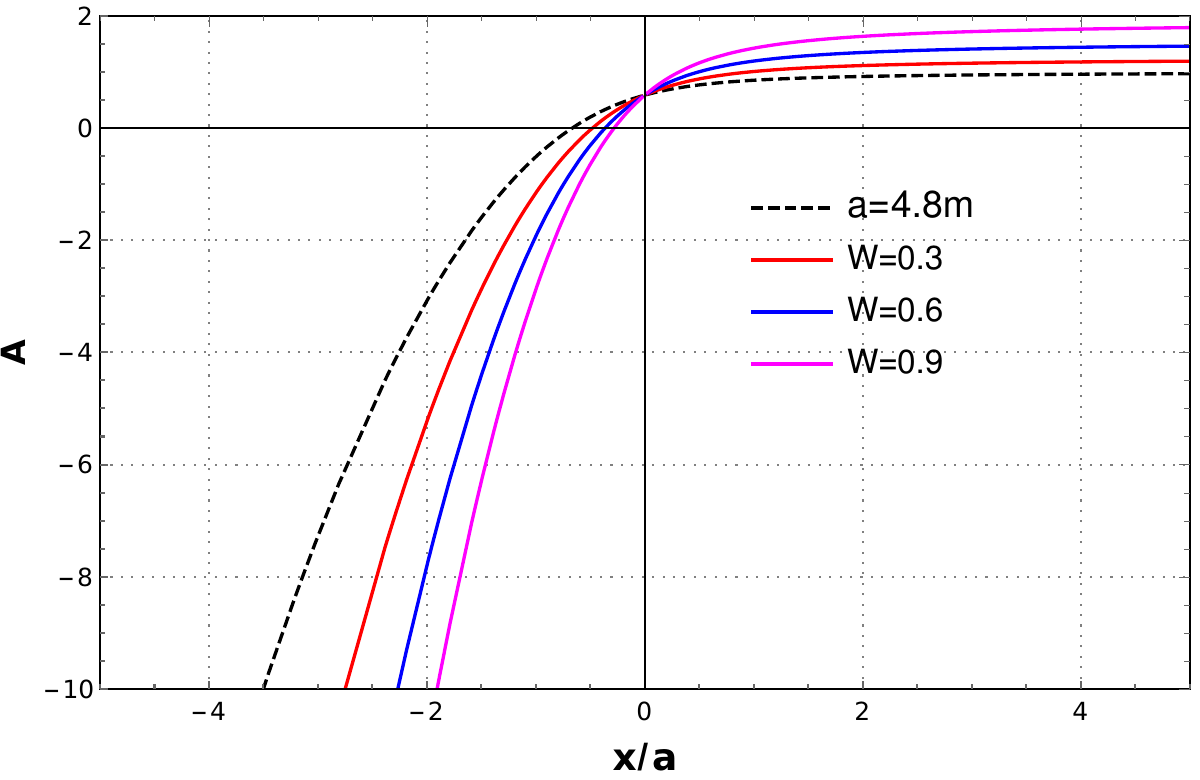}}}}
	\mbox{
	\subfigure[]{\label{K1MOD1}
	{\includegraphics[width=0.45\linewidth]{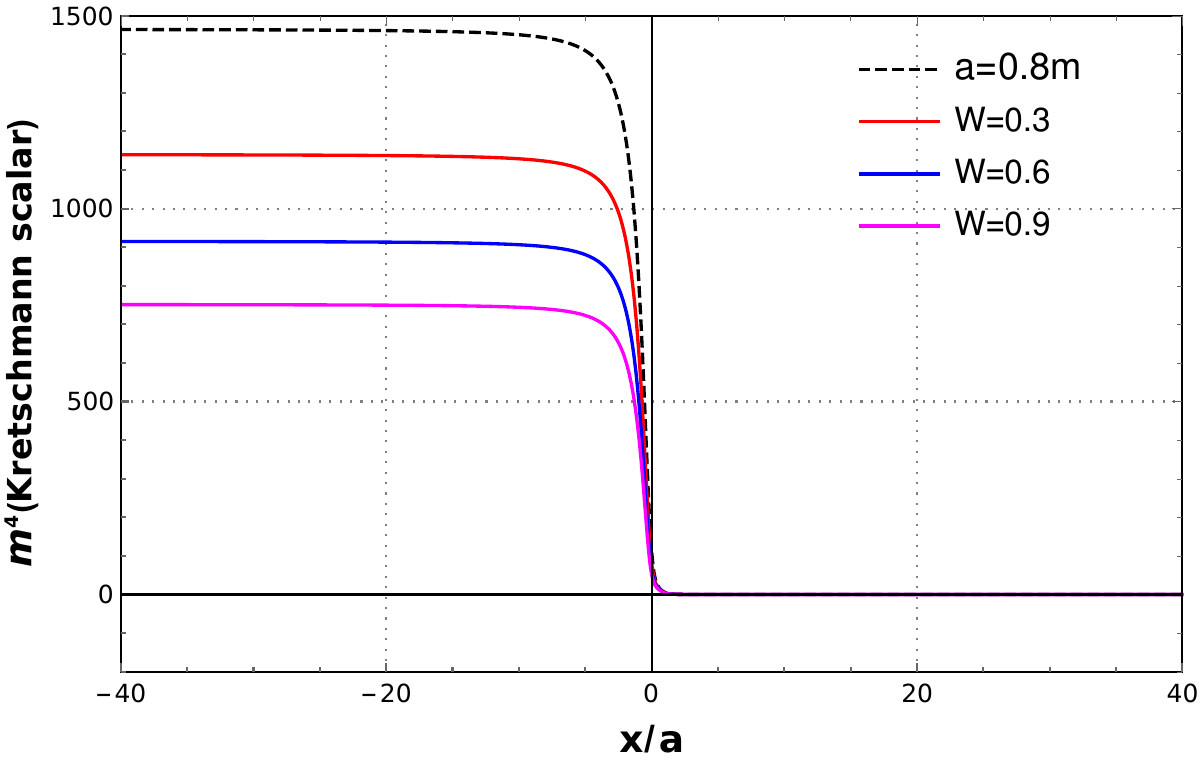}}} \qquad
	\subfigure[]{\label{K2MOD1}
	{\includegraphics[width=0.45\linewidth]{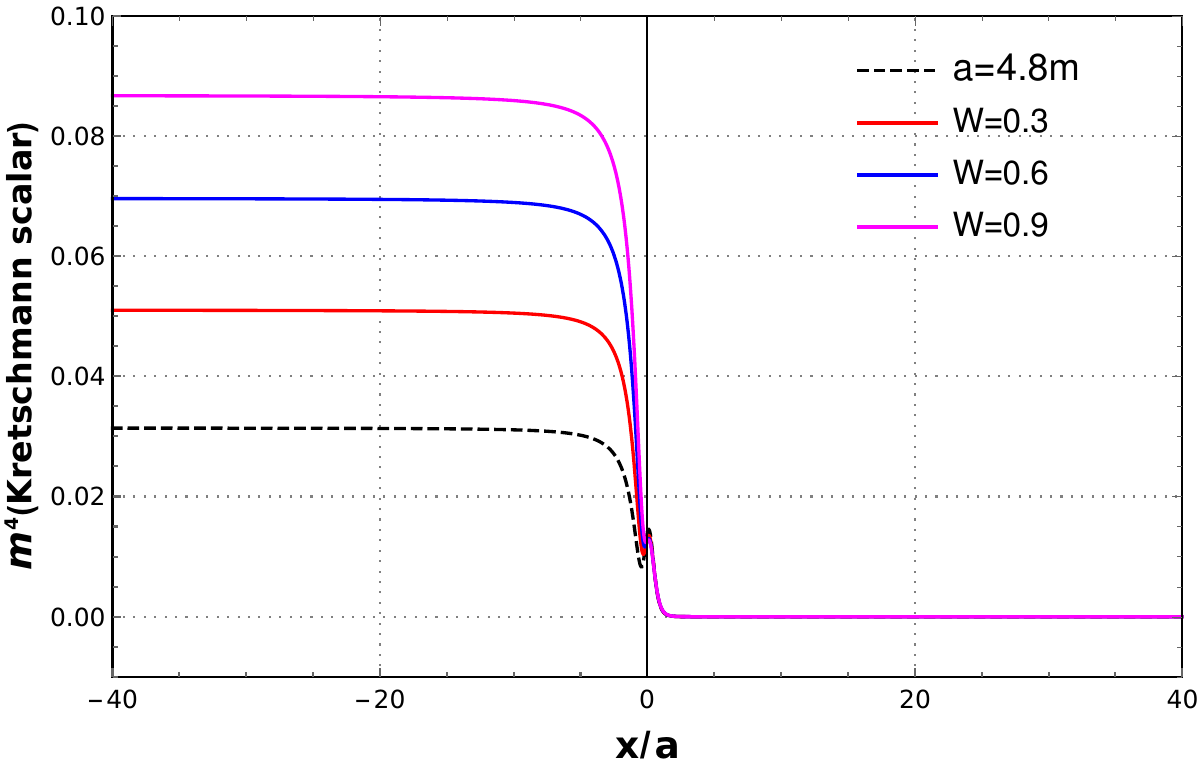}}}
}
\caption{Graphical representation for the metric function (\ref{eq15}) and the Kretschmann scalar (\ref{eq19}) for a fixed throat radius value inside and outside the horizon and varying the LSV parameter. The dotted black curve represents $W=0$. In (a) and (c), we choose values of a such that there are event horizons in the positive $x$ region, while in (b) and (d), we choose values of a such that there are no event horizons for $x>0$.}
\label{FIG1}
\end{figure}

In Fig. \ref{FIG1}, we have a graphical representation of the metric function Eq. (\ref{eq15}) for a throat value inside \ref{A1MOD1} and another outside \ref{A2MOD1} of the event horizon, and then we perform the variation of the LSV parameter. We can observe in Figs. \ref{A1MOD1} and \ref{A2MOD1} that in the limit where $x\to\infty$, the metric function has to deviate more and more from the flat solution as we increase the values of the violation parameter. On the other hand, in the limit where $x\to{-\infty}$, as we increase the violation parameter, the cosmological constant has to assume values of smaller magnitudes. This analysis can be complemented by observing the Kretschmann scalar in Figs. \ref{K1MOD1} and \ref{K2MOD1}. Although the effects of violating Lorentz symmetry in the limit where $x\to\infty$ exist, they are not sufficient to destroy the flatness of spacetime, causing the Kretschmann scalar to tend to zero. For the limit of $x\to{-\infty}$, the Kretschmann scale tends to assume smaller values when we increase the values of the violation parameter inside the horizon, and larger values for the region outside the horizon.

\begin{figure}[htb!]
\centering  
	\mbox{
	\subfigure[]{\label{PHI1MOD1}
	{\includegraphics[width=0.45\linewidth]{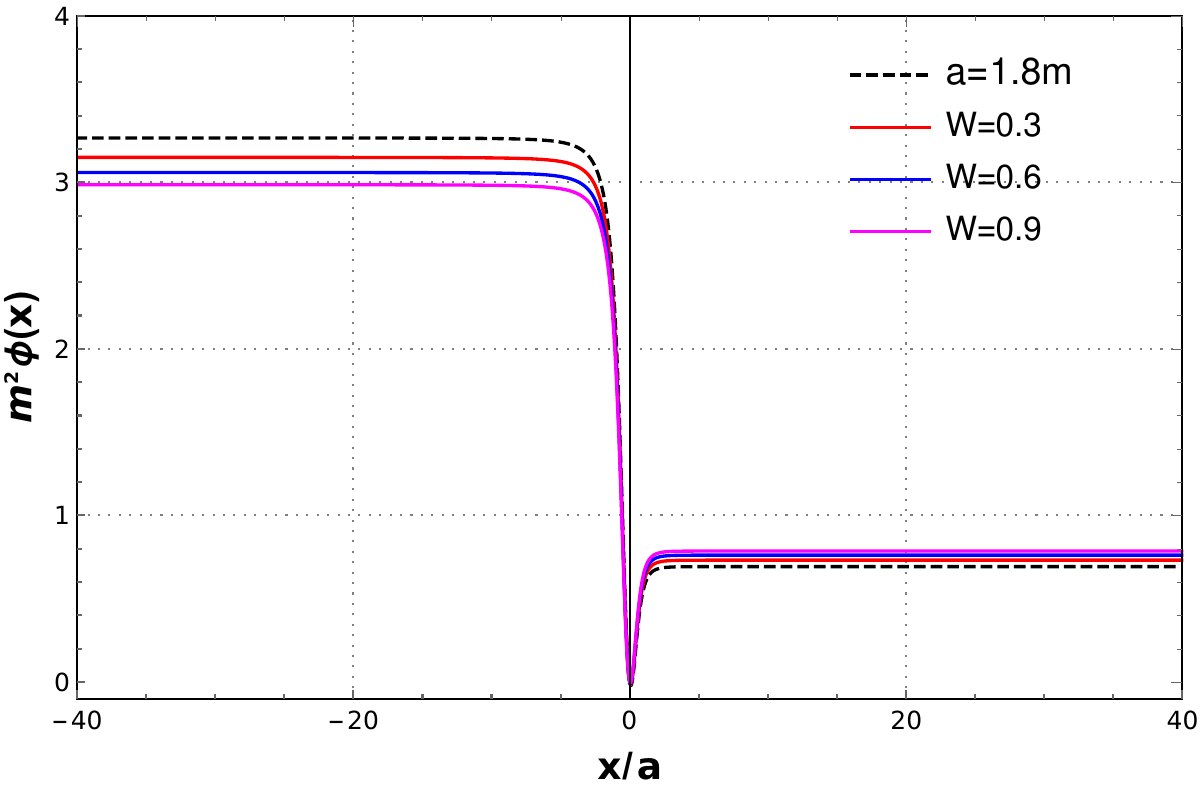}}}\qquad
	\subfigure[]{\label{PHI2MOD1}
	{\includegraphics[width=0.45\linewidth]{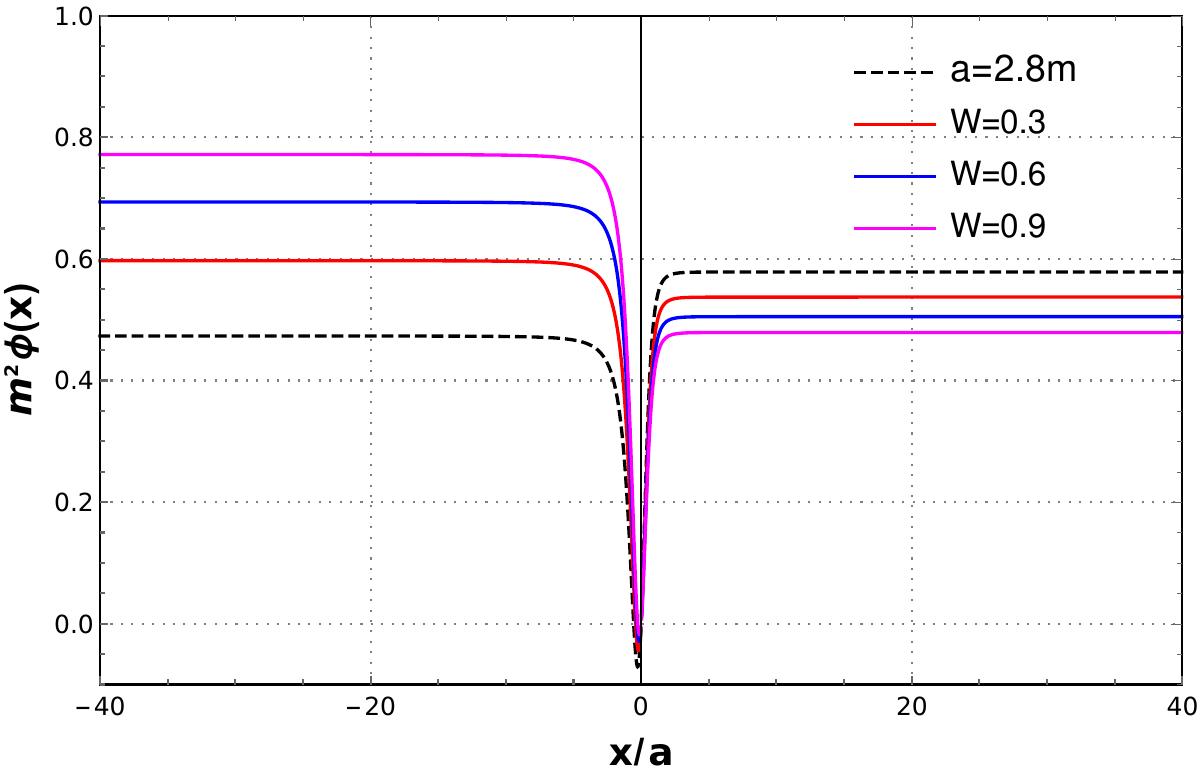}}}}
	\mbox{
	\subfigure[]{\label{POT1MO1}
	{\includegraphics[width=0.45\linewidth]{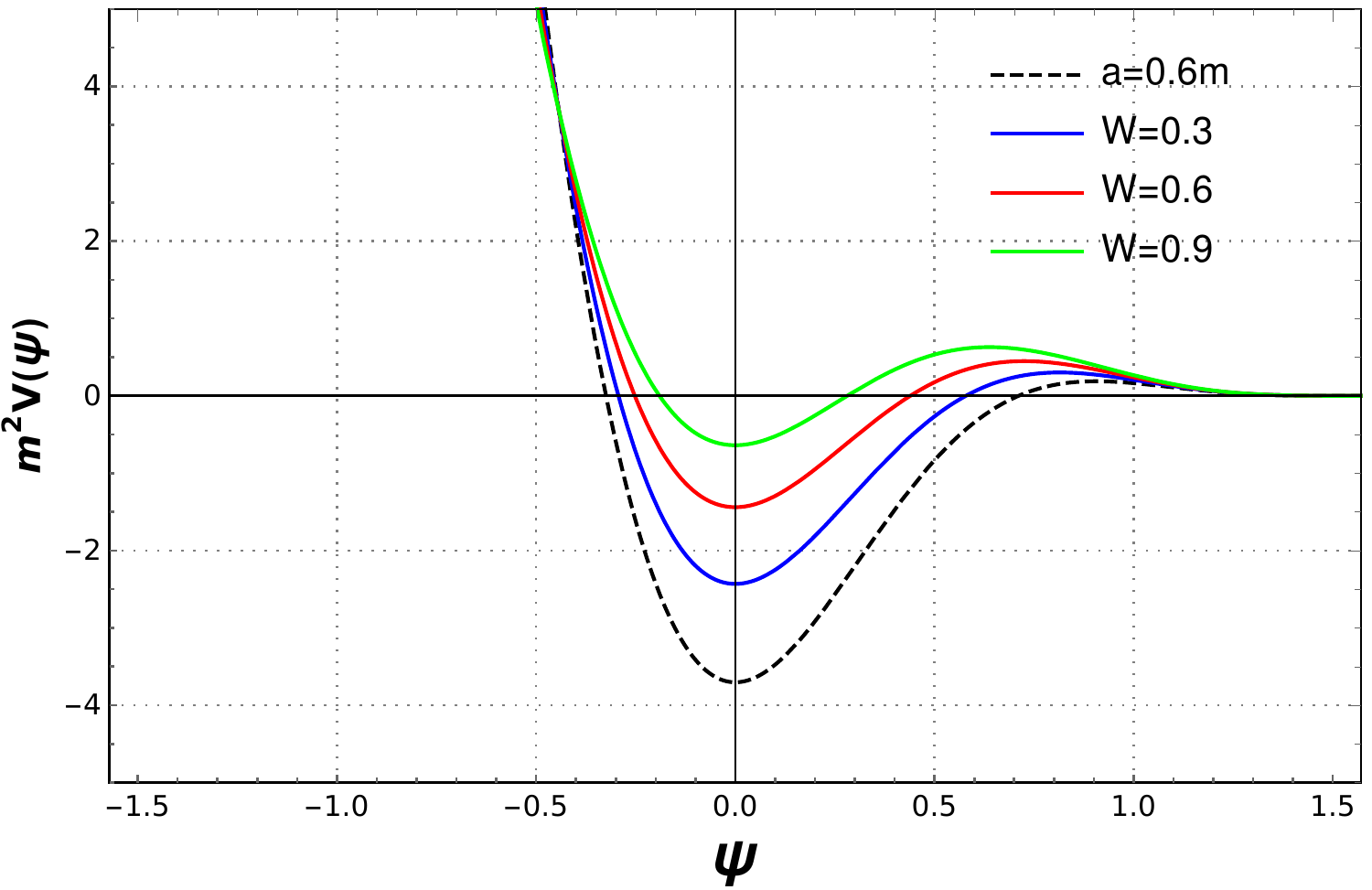}}} \qquad
	\subfigure[]{\label{POT2MO1}
	{\includegraphics[width=0.45\linewidth]{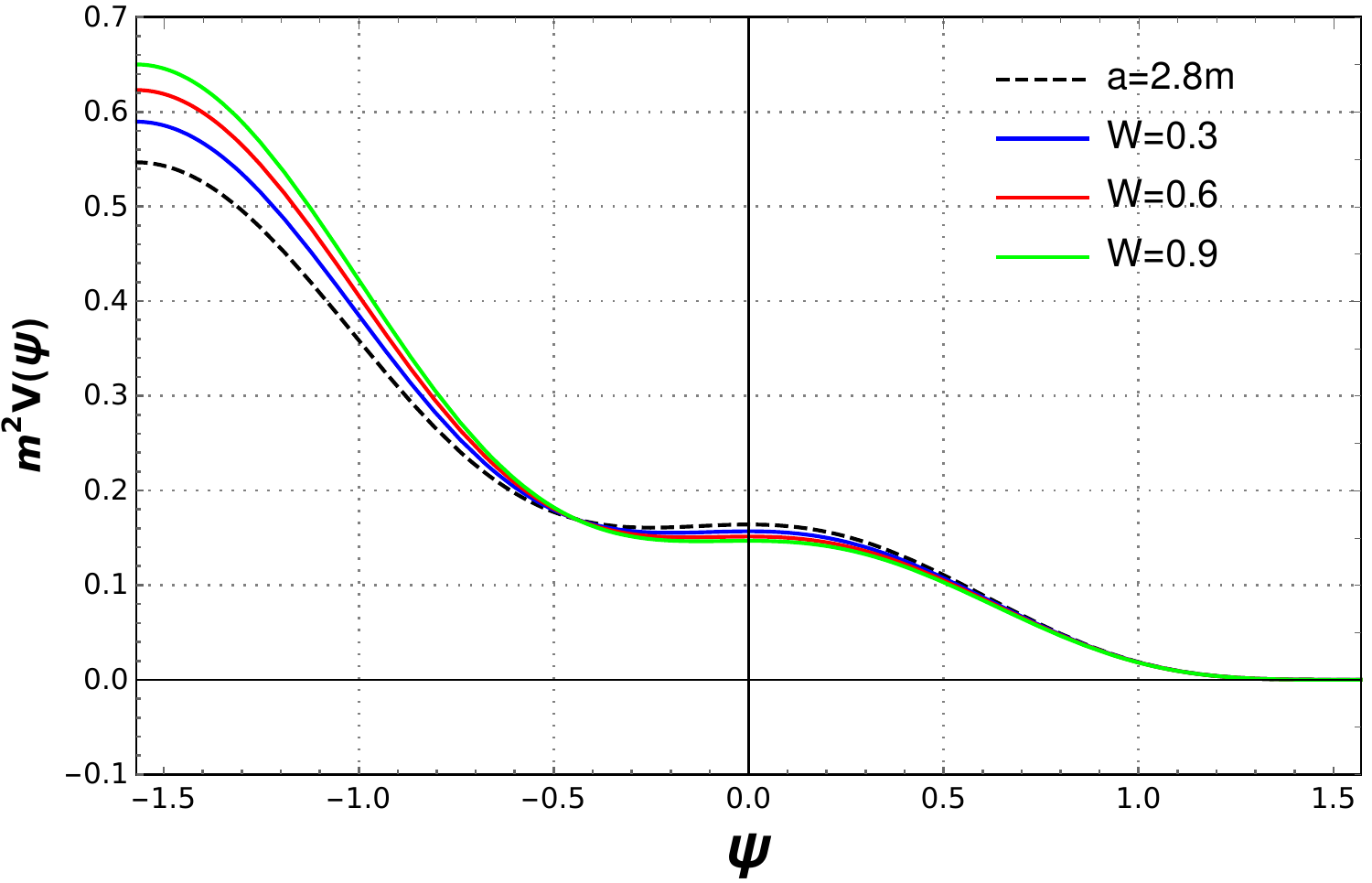}}}
}
\caption{Graphical representation for the scalar field (\ref{eq20}) and the potential (\ref{eq23}) for a fixed throat radius value inside and outside the horizon and varying the LSV parameter. We also fix $F_0=1$.The dotted black curve represents $W=0$. In (a) and (c), we depict the scalar
field and its associated potential, respectively, for the cases in which there are event horizons for $x>0$, while in (b) and (d) we show the
same functions for the cases without event horizons in $x>0$.}
\label{FIG2}
\end{figure}

In Fig. \ref{FIG2} we have a graphical representation for the scalar field Eq. (\ref{eq20}) and for the potential Eq. (\ref{eq23}), where we fix a value for the throat radius inside and another outside the event horizon and vary some values for the LSV parameter. Regarding the scalar field, for a fixed throat radius outside the event horizon in the limit where $x\to\infty$ Fig. \ref{PHI2MOD1}, as we increase the value of the violation parameter, the scalar field tends to zero more quickly. For a fixed throat radius inside the horizon in Fig. \ref{PHI1MOD1}, in this same limit, the scalar field tends to increase in amplitude as we increase the violation parameter. Likewise, the potential Fig. \ref{POT2MO1} for a radius of the throat fixed outside the horizon tends to zero when $x\to\infty$ and, as we increase the violation parameter, it tends to assume values of increasingly greater magnitudes in the limit of $x\to{-\infty}$. For a throat radius fixed inside the horizon in Fig. \ref{POT1MO1}, we can see that as we increase the value of the violation parameter, the potential minimum tends to be positively shifted at the origin and to assume values of smaller magnitudes at $x\to{-\infty}$. We saw in previous works \cite{CDJM1,CDJM2} that there is a radius regime internal to the event horizon that may indicate some stability when subjected to radial perturbations, and the presence of a LSV parameter may hinder this possibility since the potential minimum tends to be fixed positive.

\section{General solution for the second model} \label{sec4}

In this section, we are interested in investigating a new black-bounce solution in the $k$-essence context subjected to LSV effects, now considering as a starting point the introduction of a new area function $\Sigma^2_1(x)=\sqrt{x^4+d^4}$. Modifications introduced in the area function for creating new black-bounce solutions were explored in Ref. \cite{MM1}. Following the same construction process as in the previous section, we will consider the new area function above in the differential equation (\ref{eq11}) and then look for the corresponding metric function. We have the following general solution:
\begin{eqnarray}\label{eq24}
A(x)&=&\frac{\sqrt{d^4+x^4} \left(4 c_1 d^3+\left(\sqrt{2} c_2+4 d \Pi \right) \tan ^{-1}\left(\frac{\sqrt{2} x}{d}+1\right)+\left(4 d \Pi -\sqrt{2} c_2\right) \tan ^{-1}\left(1-\frac{\sqrt{2} x}{d}\right)\right)}{4 d^3} \nonumber \\
&+& \frac{c_2 \sqrt{d^4+x^4} \left(\log \left(d^2+\sqrt{2} d x+x^2\right)-\log \left(d^2-\sqrt{2} d x+x^2\right)\right)}{4 \sqrt{2} d^3}.
\end{eqnarray}

For our purposes, we will assume that the above solution is asymptotically flat in the limit $x\to\infty$ and the LSV factor is off $\Pi=1$, and we recover the Simpson-Visser limit at the origin. Thus, the above integration constants are related as follows, $c_1=- \frac{\pi{c_2}}{2\sqrt{2}d^3}$ and $c_2= \frac{d}{\sqrt{2}}\left[2\Pi-\frac{4}{\pi}\left(1-\frac{2m}{d}\right)\right]$. Thus, the general solution is as follows:
\begin{eqnarray}\label{eq25}
    A(x)&=&   \frac{\sqrt{d^4+x^4} \left(\pi  (2 d-4 m-\pi  d \Pi)+(3 \pi  d \Pi -2 d+4 m) \tan ^{-1}\left(\frac{\sqrt{2} x}{d}+1\right)+(d (\pi  \Pi +2)-4 m) \tan ^{-1}\left(1-\frac{\sqrt{2} x}{d}\right)\right)}{2 \pi  d^3} \nonumber \\
    &-& \frac{\sqrt{d^4+x^4} (d (\pi  \Pi -2)+4 m) \left(\log \left(d^2-\sqrt{2} d x+x^2\right)-\log \left(d^2+\sqrt{2} d x+x^2\right)\right)}{4 \pi  d^3}.
\end{eqnarray}

\begin{figure}[htb!]
\centering  
	\mbox{
	\subfigure[]{\label{A1MOD2}
	{\includegraphics[width=0.45\linewidth]{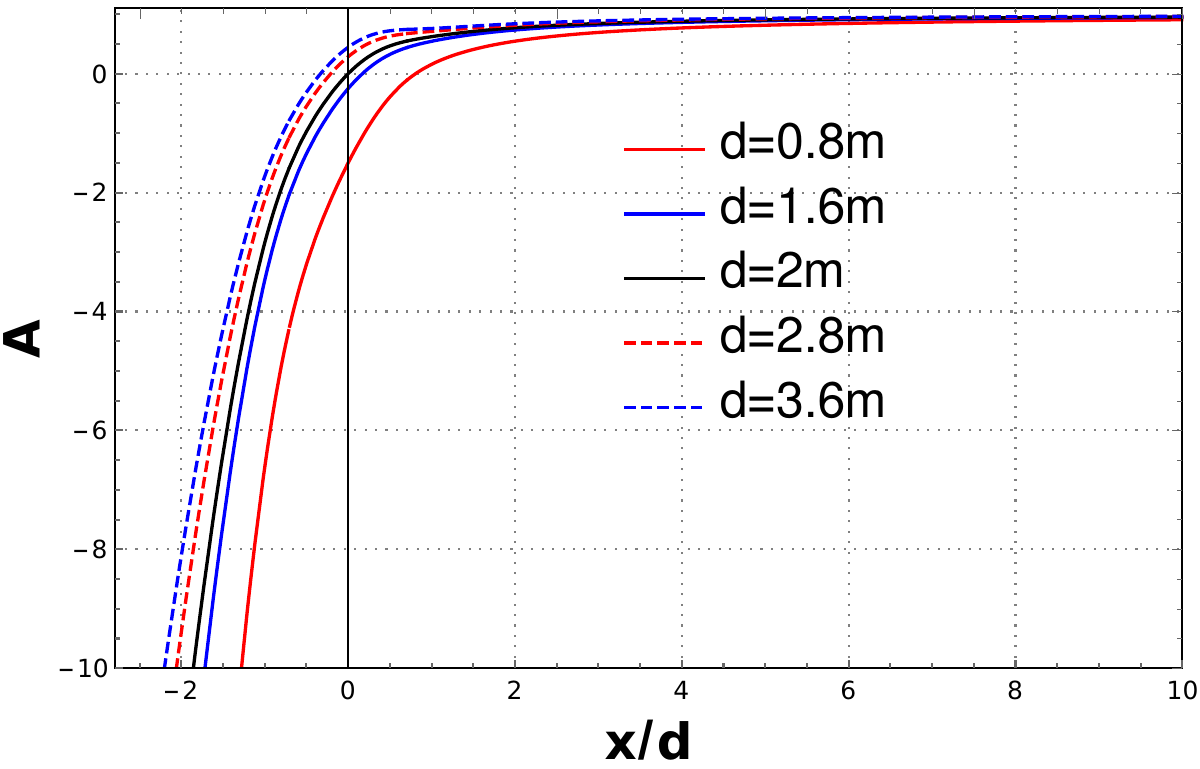}}}\qquad
	\subfigure[]{\label{K1MOD2}
	{\includegraphics[width=0.45\linewidth]{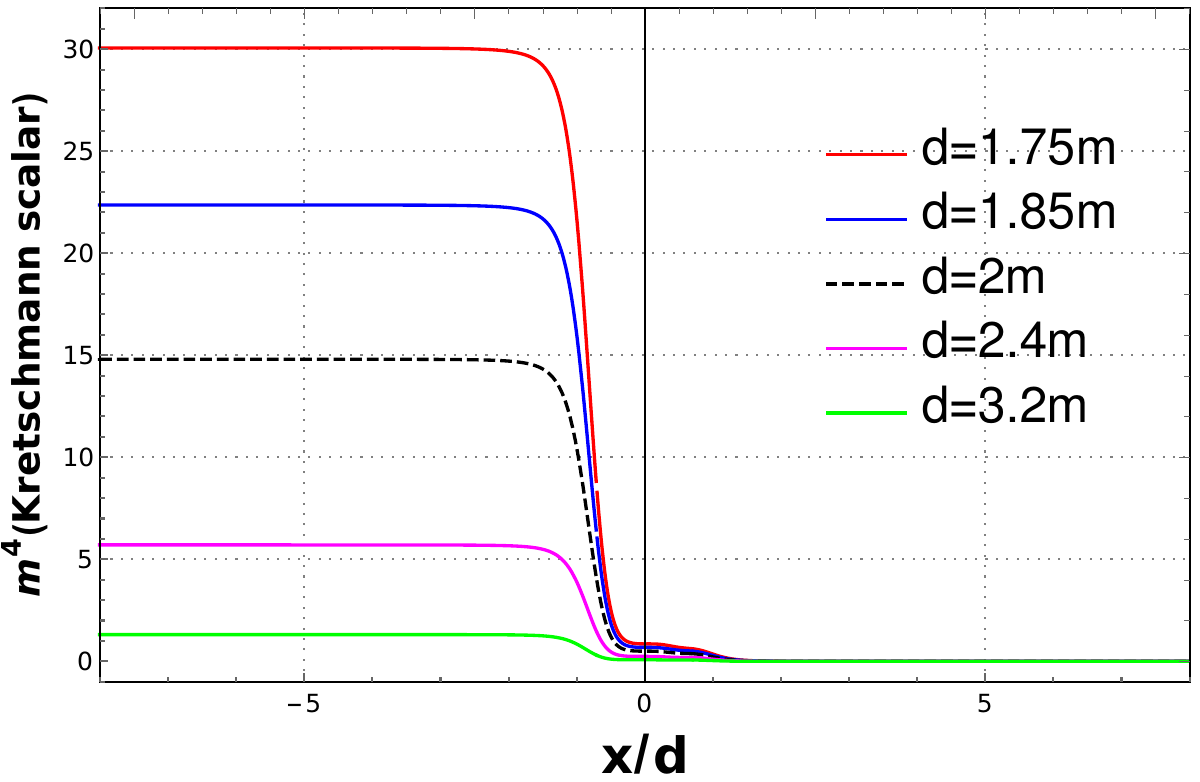}}}}
\caption{ Graphical representation for the metric function (\ref{eq25}) and the Kretschmann scalar for throat radii inside and outside the event horizon. In this context, the effects of LSV are turned off, $\Pi=1$. In (a) we have the metric function $A(x)$ and in (b) the Kretschmann scalar.}
\label{FIG3}
\end{figure}

 The above metric function has as an asymptotic limit $A(x\to\infty)=\Pi$, and at the origin, it behaves like the Simpson-Visser spacetime $A(x\to{0})= 1-\frac{2m}{d}$. At the other asymptotic $x\to{-\infty}$, the above metric function behaves like Schwarzschild-de-Sitter, and this behavior can be visualized when we perform the expansion of Eq. (\ref{eq25}):
 \begin{eqnarray}\label{eq26}
     A(x) \approx \Pi - \frac{\sqrt{2} (d (\pi  \Pi -2)+4 m)}{3 \pi  x} -\frac{x^2 (d (\pi  \Pi -2)+4 m)}{d^3}.
 \end{eqnarray} Note that for such behavior to occur, there is a restriction between the radius of the wormhole's throat and its mass $4m+d(\pi\Pi-2)>0$. In Fig. \ref{FIG3}, we have a graphical representation for the metric function Eq. (\ref{eq25}) as well as the Kretschmann scalar for some throat radii inside and outside the event horizon. In Fig. \ref{K1MOD2}, we can observe that the Kretschmann scalar tends to zero when $x\to\infty$, as expected for an asymptotically flat spacetime, and in the limit of $x\to{-\infty}$ it tends to assume increasingly positive values, as we consider throat radii increasingly internal to the event horizon.

\begin{figure}[htb!]
\centering  
	\mbox{
	\subfigure[]{\label{A2MOD2}
	{\includegraphics[width=0.45\linewidth]{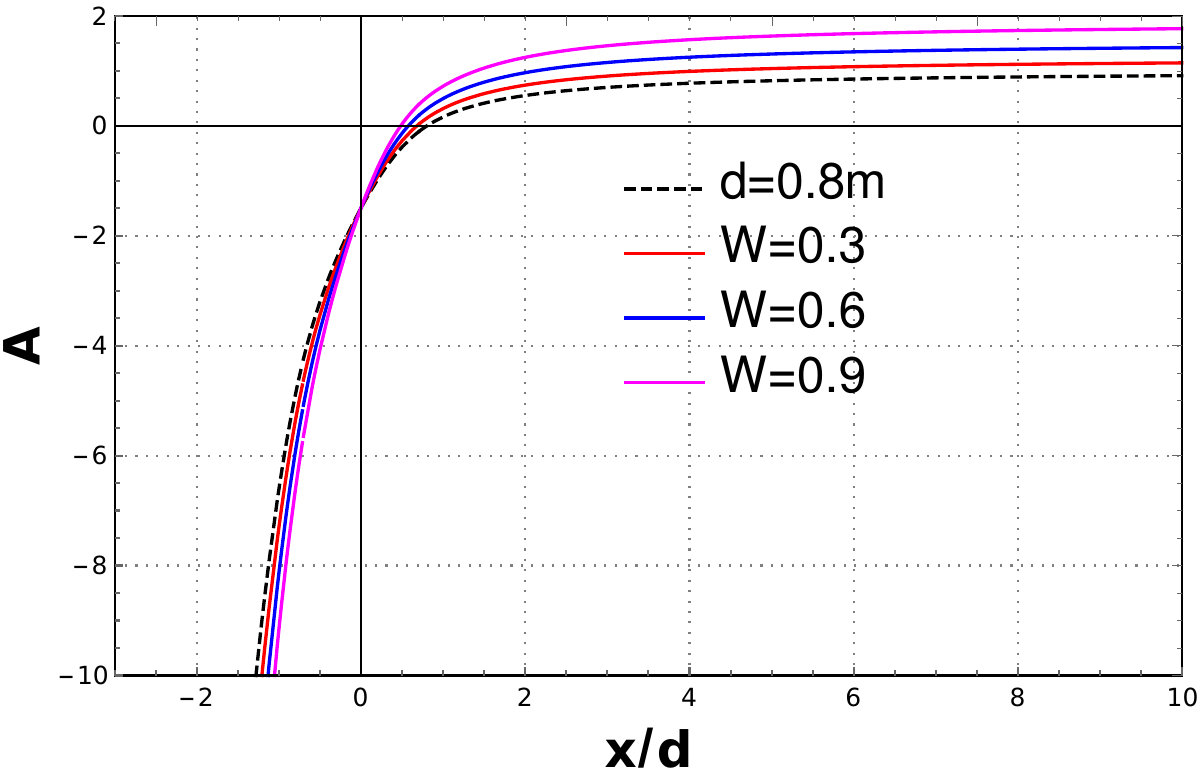}}}\qquad
	\subfigure[]{\label{A3MOD2}
	{\includegraphics[width=0.45\linewidth]{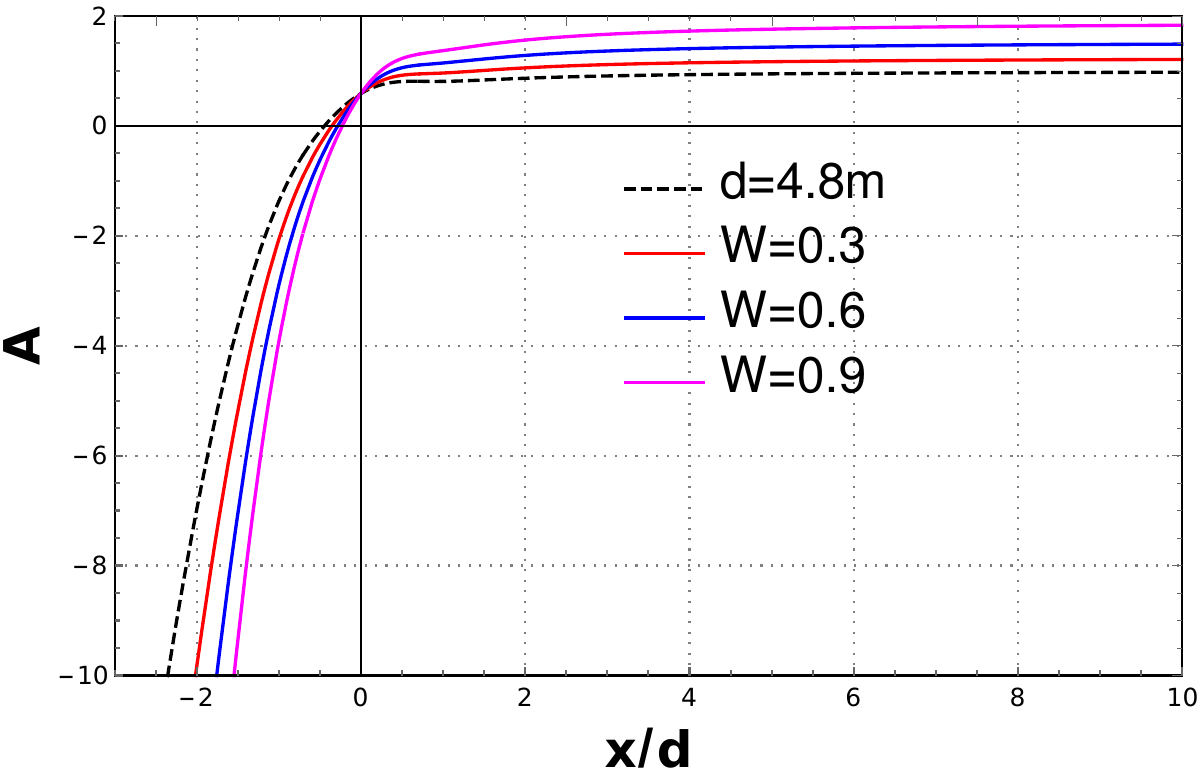}}}}
	\mbox{
	\subfigure[]{\label{K2MOD2}
	{\includegraphics[width=0.45\linewidth]{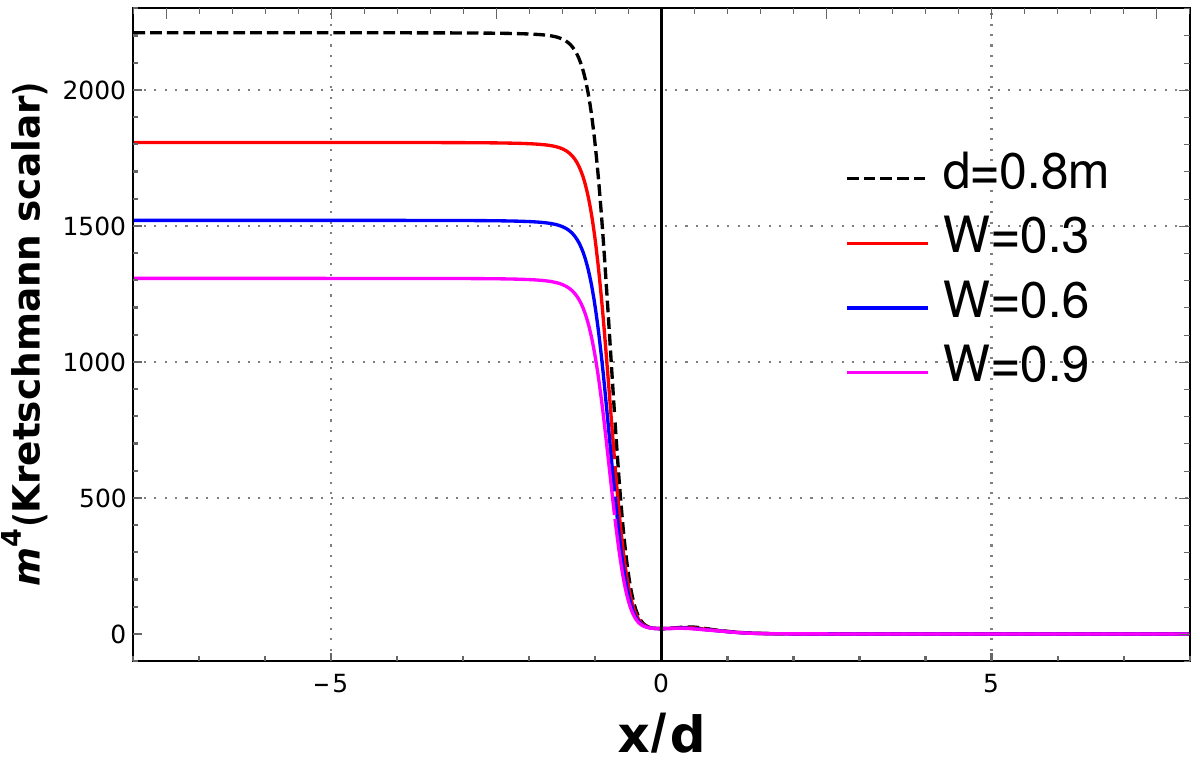}}} \qquad
	\subfigure[]{\label{K3MOD2}
	{\includegraphics[width=0.45\linewidth]{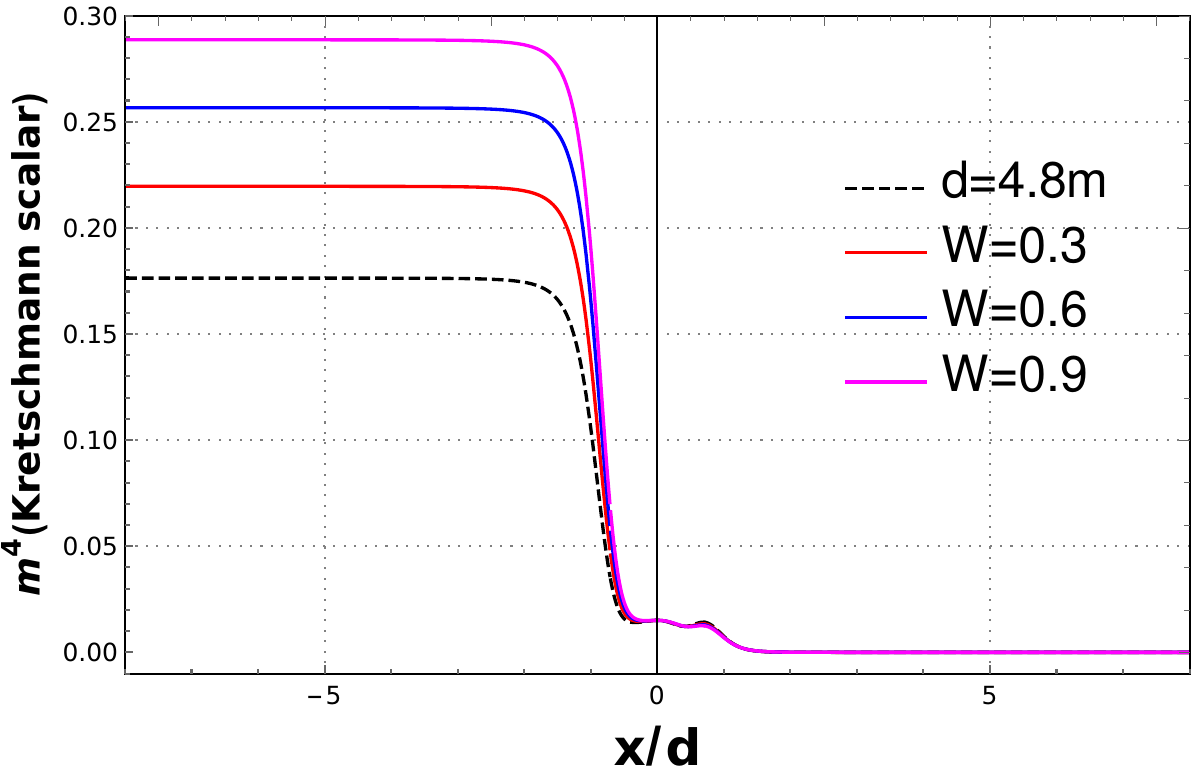}}}
}
\caption{Graphical representation for the metric function (\ref{eq25}) and the Kretschmann scalar for a fixed throat radius value inside and outside the horizon and varying the LSV parameter. The dotted black curve represents $W=0$. In (a) and (c) we have the metric functions and the Kretschmann scalar, respectively, to the cases where there is an event horizon in the region $x>0$ while in (b) and (d) we have the
same functions to the cases where there is no horizon in $x>0$.}
\label{FIG4}
\end{figure}

In Fig. \ref{FIG4}, we have a graphical representation of the metric function Eq. (\ref{eq25}) together with the Kretschmann scalar for a throat radius fixed inside and another outside the event horizon, and then we perform the variation of the LSV parameter. In Figs. \ref{A2MOD2} and \ref{K2MOD2}, we set the radius of the wormhole throat at $d=0.8m$ and then vary the LSV parameter, where we can see that in the limit of $x\to\infty$, the spacetime is still asymptotically flat but is being positively deflected by the violation parameter. On the other hand, in the limit where $x\to{-\infty}$, the LSV factor causes the spacetime to have a cosmological constant with a smaller magnitude. In Figs. \ref{A3MOD2} and \ref{K3MOD2}, we fix a throat radius outside the horizon $d=4.8m$ and vary the LSV parameter, and then we have a similar behavior in the limit of $x\to \infty$, and when $x\to{-\infty}$, spacetime tends to assume values for a cosmological constant with greater magnitudes.

\begin{figure}[htb!]
\centering  
	\mbox{
	\subfigure[]{\label{PHI1MOD2}
	{\includegraphics[width=0.45\linewidth]{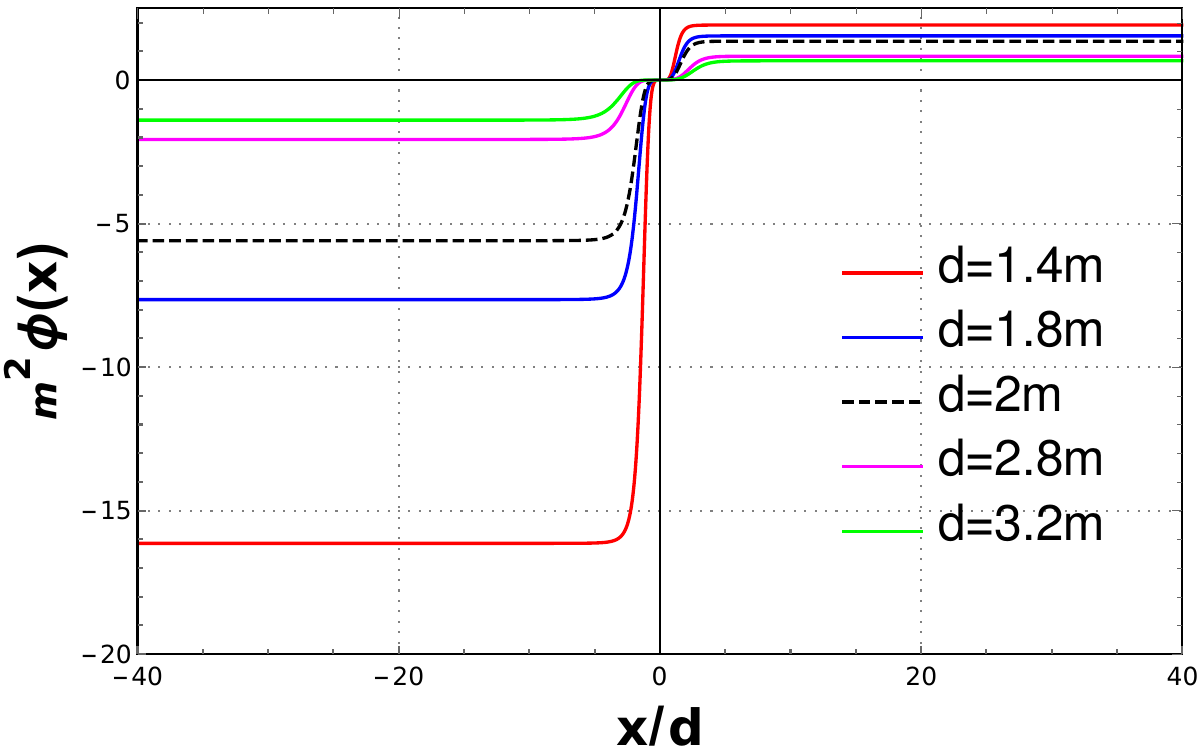}}}\qquad
	\subfigure[]{\label{POT1MOD2}
	{\includegraphics[width=0.45\linewidth]{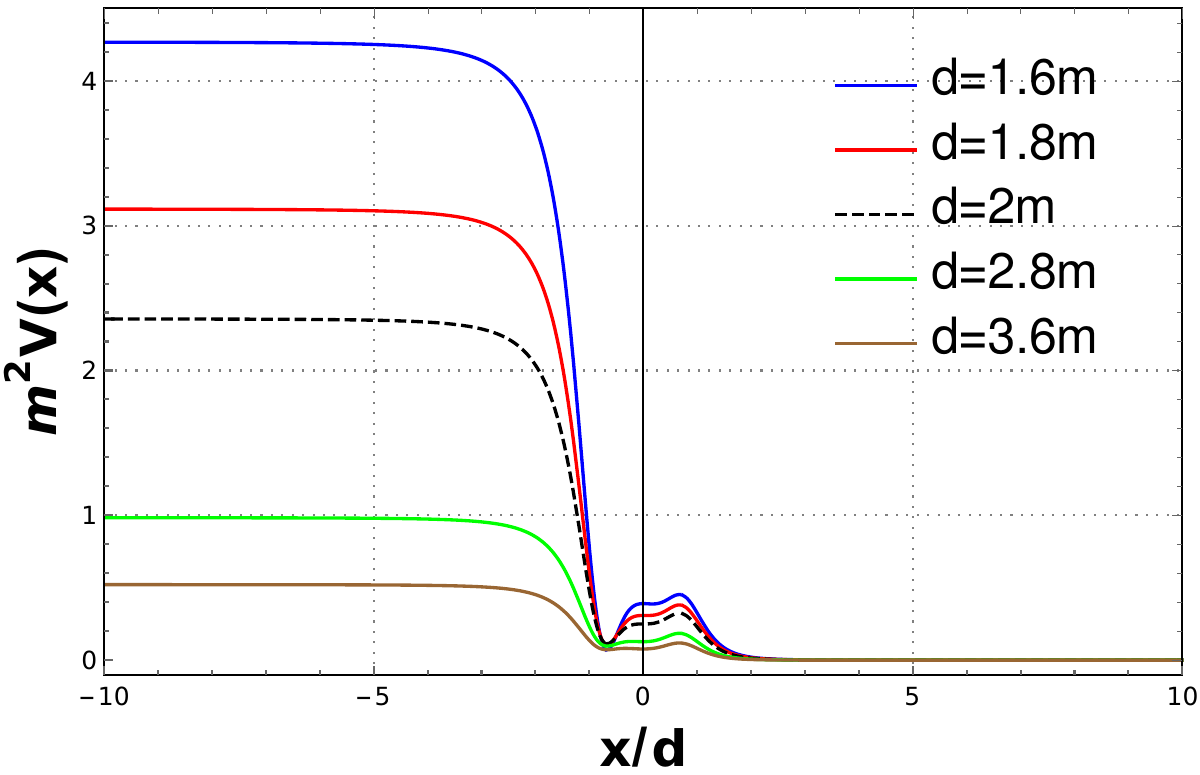}}}}
\caption{Plot for the scalar field and potential (\ref{eq26a}) for throat radii inside and outside the event horizon. In this context, the effects of LSV are turned off, $\Pi=1$, and we fix $F_0=1$. In (a) we have the scalar field $\phi(x)$ and in (b) the potential $V(x)$.}
\label{FIG5}
\end{figure}

Using the combination of Eqs. (\ref{eq10}) and (\ref{eq13}) for the $k$-essence field and the metric functions Eq. (\ref{eq25}) and $\Sigma^2_1(x)$, we obtain the expression $2A{\Sigma}''_1/\Sigma_1=\Pi{XF_X}$. For the configuration of the $k$-essence field $n=1/3$, we can get the scalar field. However, we perform the numerical treatment discussed in the following because of algebraic difficulties.

Similarly, using Eq. (\ref{eq13}) for the metric functions Eq. (\ref{eq25}) and $\Sigma^2_1(x)$ with the help of the expression $2A{\Sigma}''_1/\Sigma_1=\Pi{XF_X}$, we then obtain the potential expression below, for the configuration $n=1/3$,
\begin{eqnarray}\label{eq26a}
V(\phi(x)) &=& \frac{2 \pi  d^7 \Pi -3 \pi  d^4 x^2 (d (\pi  \Pi -2)+4 m)-2 \sqrt{2} d^3 x^3 (d (\pi  \Pi -2)+4 m)+6 \pi  d^3 \Pi  x^4+3 \pi  x^6 (d (\pi  \Pi -2)+4 m)}{2 \pi  d^3 \Pi  \left(d^4+x^4\right)^{3/2}} \nonumber \\
&+& \frac{3 x^2 \left(d^4-x^4\right) \left((d (3 \pi  \Pi -2)+4 m) \tan ^{-1}\left(\frac{\sqrt{2} x}{d}+1\right)+(d (\pi  \Pi +2)-4 m) \tan ^{-1}\left(1-\frac{\sqrt{2} x}{d}\right)\right)}{2 \pi  d^3 \Pi  \left(d^4+x^4\right)^{3/2}} \nonumber \\
&+& \frac{3 x^2 \left(x^4-d^4\right) (d (\pi  \Pi -2)+4 m) \left(\log \left(d^2-\sqrt{2} d x+x^2\right)-\log \left(d^2+\sqrt{2} d x+x^2\right)\right)}{4 \pi  d^3 \Pi  \left(d^4+x^4\right)^{3/2}}.
\end{eqnarray}

In Fig. \ref{FIG5}, we have a graphical representation of the scalar field and the potential associated with it Eq. (\ref{eq26a}) for some values of throat radii inside and outside the event horizon, taking into account that there are no effects of LSV, $\Pi=1$. In Fig. \ref{PHI1MOD2}, we have the numerical solution for the scalar field, considering as a reference the throat radius, located at the horizon $d=2m$, and then we vary it for the internal and external regions. In the limit $x\to\infty$, the scalar field tends to assume positive values and can increase in magnitude for throat radii that are more internal to the horizon and decrease to zero for radii that are increasingly external to the horizon.

In Fig. \ref{POT1MOD2}, we have the profiles of the potential Eq. (\ref{eq26a}), with the black dotted curve for the radius of the throat located on the horizon $d=2m$, and then we vary the radius for part of the internal and external regions. In the limit of $x\to\infty$, the potential assumes constant and positive values and, as we get further outside the horizon, it tends to zero. In the limit in which $x\to{-\infty}$, it tends to assume positive values of larger magnitudes for throat radii that are more internal to the horizon and smaller values for more external radii. 

The potential minimum, located to the left of the origin in Fig. \ref{POT1MOD2}, tends to be shifted to the negative region, which may indicate the possibility of some type of stability in this regime when the scalar field is under radial perturbations.

\section{General relations for energy conditions}\label{sec5}

The analysis of energy conditions, as presented in \cite{LIVRO1}, starts from Einstein's equations, obtained from Eq. (\ref{eq2}). Subsequent calculations produce the following non-vanishing components of the stress-energy tensor \cite{LIVRO2}:
\begin{eqnarray}\label{eq27}
\tensor{T}{^\mu}_{\nu}= {\rm diag}\left[\rho^\phi,-p^\phi_1,-p^\phi_2,-p^\phi_2\right],
\end{eqnarray} where the scalar field energy density is denoted by $\rho^\phi$, while $p^\phi_1$ and $p^\phi_2$ represent the radial and tangential pressures, respectively. When using the diagonal components of the stress-energy tensor, as expressed in Eqs. (\ref{eq4}) and (\ref{eq5}), for the $k$-essence configuration $n=1/3$ and the scalar potential Eq. (\ref{eq21a}), we get
\begin{eqnarray}\label{eq28}
\rho^\phi&=& -\frac{F_0}{2}\left[-\eta{\frac{A}{\Pi}{\left(\phi'\right)}^2}\right]^\frac{1}{3} + V\left(x\right)= -\frac{3{A{\Sigma}''}}{\Pi\Sigma} +V\left(x\right), \\\label{eq29} 
p^\phi_{1}&=& - T^{1}_{1} =  \frac{{A{\Sigma}''}}{\Pi\Sigma} -V\left(x\right), \\\label{eq30} 
p^\phi_{2}&=&  - T^{2}_{2}= - T^{0}_{0}=- \rho^\phi= \frac{3{A{\Sigma}''}}{\Pi\Sigma} - V\left(x\right).
\end{eqnarray}

The diagonal components of the stress-energy tensor, as defined above, are applicable solely outside the horizon, where $A>0$, with the metric signature $(+,-,-,-)$. In this domain, the coordinate $t$ is timelike, and the coordinate $x$ is spacelike.

Within the event horizon, where $A<0$, the coordinate $t$ becomes spacelike and $x$ becomes timelike. Consequently, the metric signature transforms to $(-,+,-,-)$, thereby reversing the roles of the coordinates. The stress-energy tensor components must then be reformulated as
\begin{eqnarray}\label{eq31}
\tensor{T}{^\mu}_{\nu}= {\rm diag}\left[-p^\phi_1,\rho^\phi,-p^\phi_2,-p^\phi_2\right],
\end{eqnarray} and, therefore, the equations for energy density, radial pressure, and tangential pressure must be rewritten as
\begin{eqnarray}\label{eq32}
\rho^\phi&=& -\frac{{A{\Sigma}''}}{\Pi\Sigma} +V\left(x\right),\\\label{eq33} 
p^\phi_{1}&=&  \frac{3{A{\Sigma}''}}{\Pi\Sigma} - V\left(x\right),\\\label{eq34}
p^\phi_{2}&=& - T^{2}_{2}= - T^{0}_{0}=-\rho^\phi=-\left(-p^\phi_{1}\right)=\frac{3{A{\Sigma}''}}{\Pi\Sigma} -V\left(x\right).
\end{eqnarray} 

Energy conditions are typically expressed as inequalities involving energy density and pressure components \cite{LVS}
\begin{eqnarray}\label{eq35}
NEC_{1,2}&=&WEC_{1,2}=SEC_{1,2} \Longleftrightarrow \rho^\phi + p^\phi_{\left(1,2\right)} \geq 0, \\\label{eq36}
SEC_3 &\Longleftrightarrow & \rho^\phi + p^\phi_{1} + 2p^\phi_{2} \geq 0, \\\label{eq37}
DEC_{1,2} &\Longleftrightarrow &  \rho^\phi + p^\phi_{\left(1,2\right)} \geq 0  \qquad    \mbox{and} \qquad \rho^\phi - p^\phi_{\left(1,2\right)} \geq 0 , \\\label{eq38}
DEC_3&=&WEC_{3} \Longleftrightarrow   \rho^\phi  \geq 0 .
\end{eqnarray}

The energy conditions can be written explicitly in terms of the metric functions by substituting the components of the stress-energy tensor Eqs. (\ref{eq28}-\ref{eq30}) into the inequalities defined above. This gives the energy conditions in the timelike region outside the event horizon, where $A>0$, as
\begin{eqnarray}\label{eq39}
NEC^\phi_{1}&=&WEC^\phi_{1}=SEC^\phi_{1} \Longleftrightarrow  -\frac{2A\Sigma''}{\Pi\Sigma} \geq 0, \\\label{eq40}
NEC^\phi_{2}&=&WEC^\phi_{2}=SEC^\phi_{2} \Longleftrightarrow  0, \\\label{eq41}
SEC^\phi_3 & \Longleftrightarrow &  \frac{4{\Sigma}''{A}}{\Pi\Sigma} -2V\left(x\right)  \geq 0, \\\label{eq42}
DEC^\phi_{1} & \Longleftrightarrow & -\frac{4{\Sigma}''{A}}{\Pi\Sigma}                                + 2V\left(x\right) \geq 0, \\\label{eq43}
DEC^\phi_{2} & \Longleftrightarrow & -\frac{6{\Sigma}''{A}}{\Pi\Sigma}                                + 2V\left(x\right) \geq 0, \\\label{eq44}
DEC^\phi_{3}&=&WEC^\phi_{3}  \Longleftrightarrow   -\frac{3{A{\Sigma}''}}{\Pi\Sigma} +V\left(x\right) \geq 0.
\end{eqnarray}

Similarly, the energy conditions within the horizon, where $t$ is spacelike, are derived by substituting the stress-energy tensor components from Eqs. (\ref{eq32}-\ref{eq34}) into the inequalities in Eqs. (\ref{eq35}-\ref{eq38}). This results in the energy conditions, for $A< 0$, as
\begin{eqnarray}\label{eq45}
NEC^\phi_{1}&=&WEC^\phi_{1}=SEC^\phi_{1} \Longleftrightarrow  \frac{2A\Sigma''}{\Pi\Sigma} \geq 0, \\\label{eq46}
NEC^\phi_{2}&=&WEC^\phi_{2}=SEC^\phi_{2} \Longleftrightarrow   \frac{2A\Sigma''}{\Pi\Sigma} \geq 0, \\\label{eq47}
SEC^\phi_3 & \Longleftrightarrow &  \frac{8A\Sigma''}{\Pi\Sigma} -2V\left(x\right)\geq 0, \\\label{eq48}
DEC^\phi_{1} & \Longleftrightarrow & -\frac{4A\Sigma''}{\Pi\Sigma} +2V\left(x\right) \geq 0, \\\label{eq49}
DEC^\phi_{2} & \Longleftrightarrow & -\frac{4A\Sigma''}{\Pi\Sigma} +2V\left(x\right) \geq 0, \\\label{eq50}
DEC^\phi_{3}&=&WEC^\phi_{3} \Longleftrightarrow   -\frac{A\Sigma''}{\Pi\Sigma} +V\left(x\right) \geq 0.
\end{eqnarray}

Analyzing the energy conditions defined above, it is clear that the leading null energy condition $NEC^{\phi}_1$ Eq. (\ref{eq39}) and Eq. (\ref{eq45}) is violated for throat radii located both inside and outside the event horizon and, therefore, as a direct consequence, the dominant energy condition $DEC^{\phi}_1$ Eq. (\ref{eq42}) and Eq. (\ref{eq48}) is also violated. This result is independent of the LSV parameter $\Pi$ and holds for both models discussed in this work. The secondary null energy condition $NEC^{\phi}_2$ is satisfied outside the event horizon Eq. (\ref{eq40}) and violated inside Eq. (\ref{eq46}). Logically, the dominant energy condition is also violated within the horizon Eq. (\ref{eq49}).

Only viewing the expression of the dominant energy condition $DEC^{\phi}_2$ outside the horizon Eq. (\ref{eq43}), the strong energy condition $SEC^{\phi}_3$ outside Eq. (\ref{eq41}) and inside the horizon Eq. (\ref{eq47}), and the dominant energy condition $DEC^{\phi}_3$ outside Eq. (\ref{eq44}) and inside the horizon Eq. (\ref{eq50}), it is not possible to draw any conclusions. So, a graphical analysis will be carried out later.

\subsection{Analysis of complementary energy conditions first model}\label{sec5a}

The complementary energy conditions for this first model will be performed only with the LSV parameter since the normal case where the violation effects are not considered, $\Pi=1$, has already been analyzed in Refs. \cite{CDJM1,CDJM2}.

\begin{figure}[htb!]
\centering  
	\mbox{
	\subfigure[]{\label{SEC3FMOD1}
	{\includegraphics[width=0.45\linewidth]{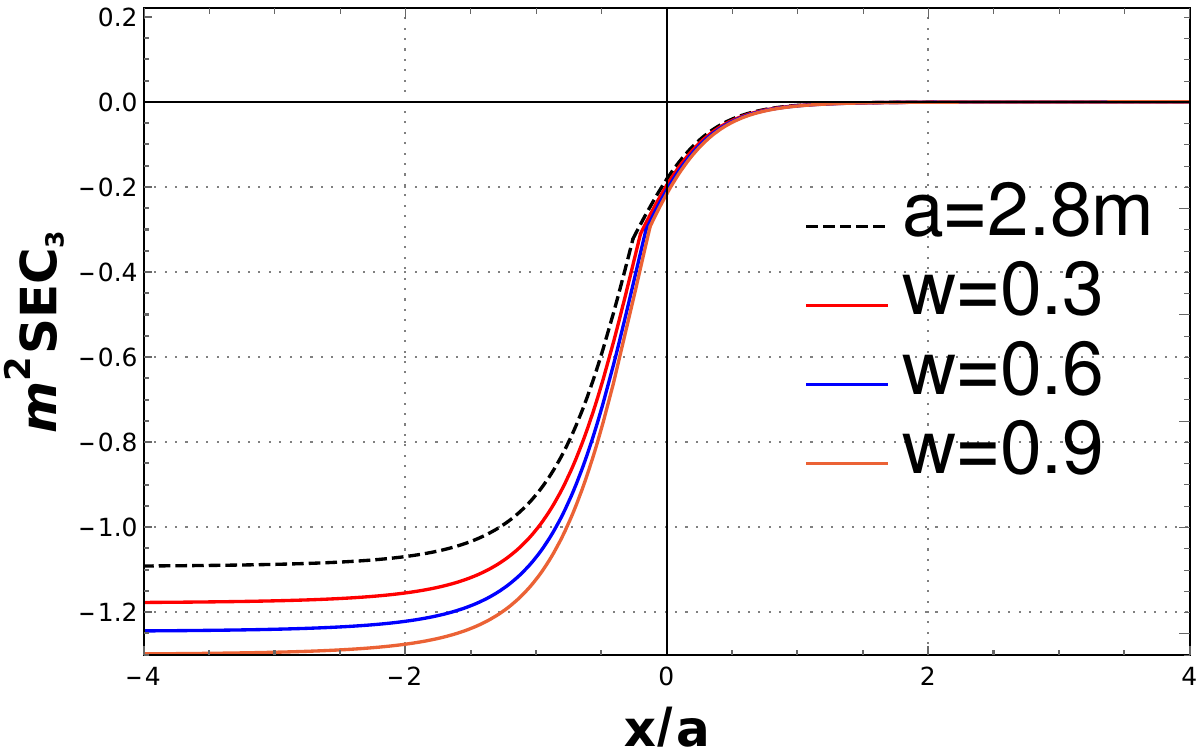}}}\qquad
	\subfigure[]{\label{SEC3DMOD1}
	{\includegraphics[width=0.45\linewidth]{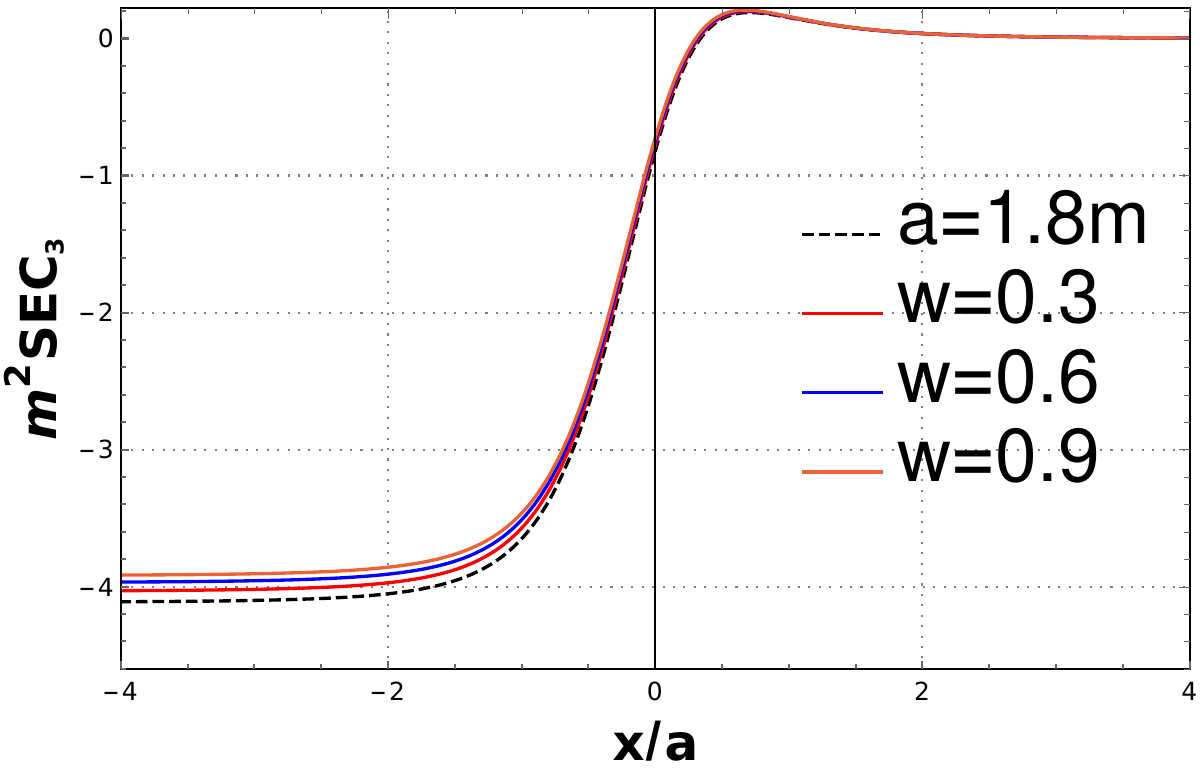}}}}
	\mbox{
	\subfigure[]{\label{DEC3FMOD1}
	{\includegraphics[width=0.45\linewidth]{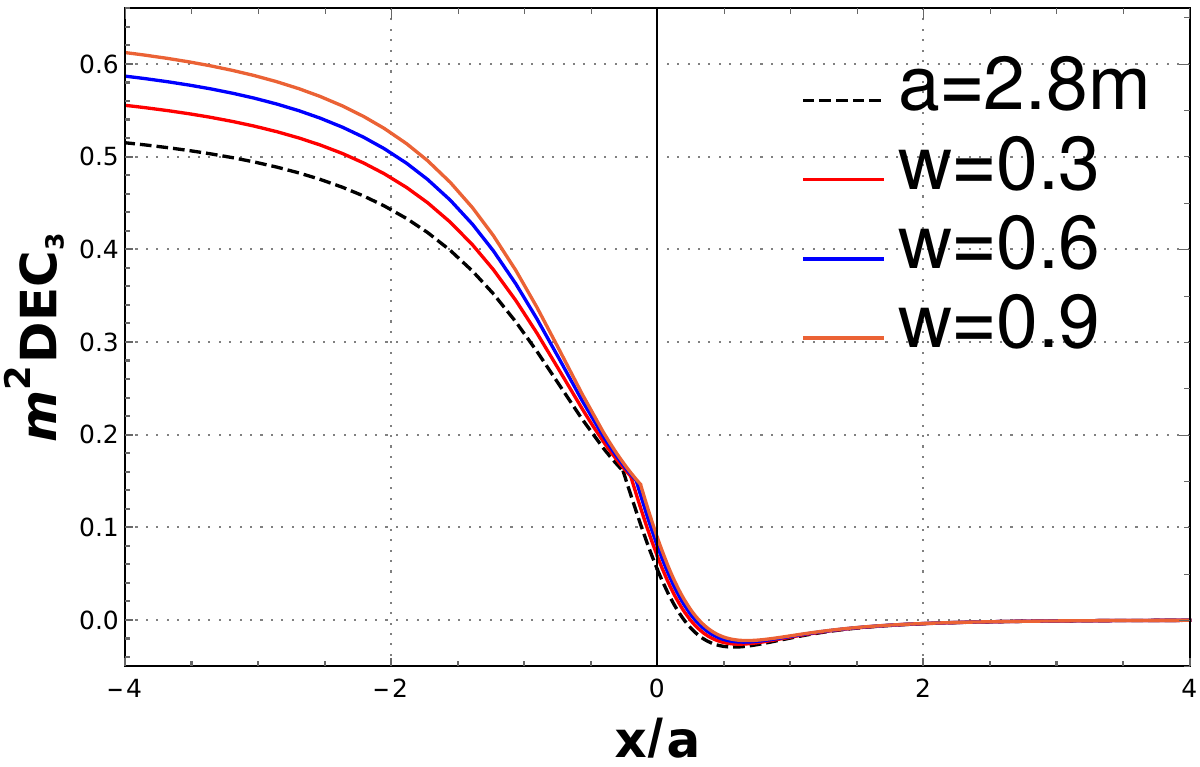}}} \qquad
	\subfigure[]{\label{DEC3DMOD1}
	{\includegraphics[width=0.45\linewidth]{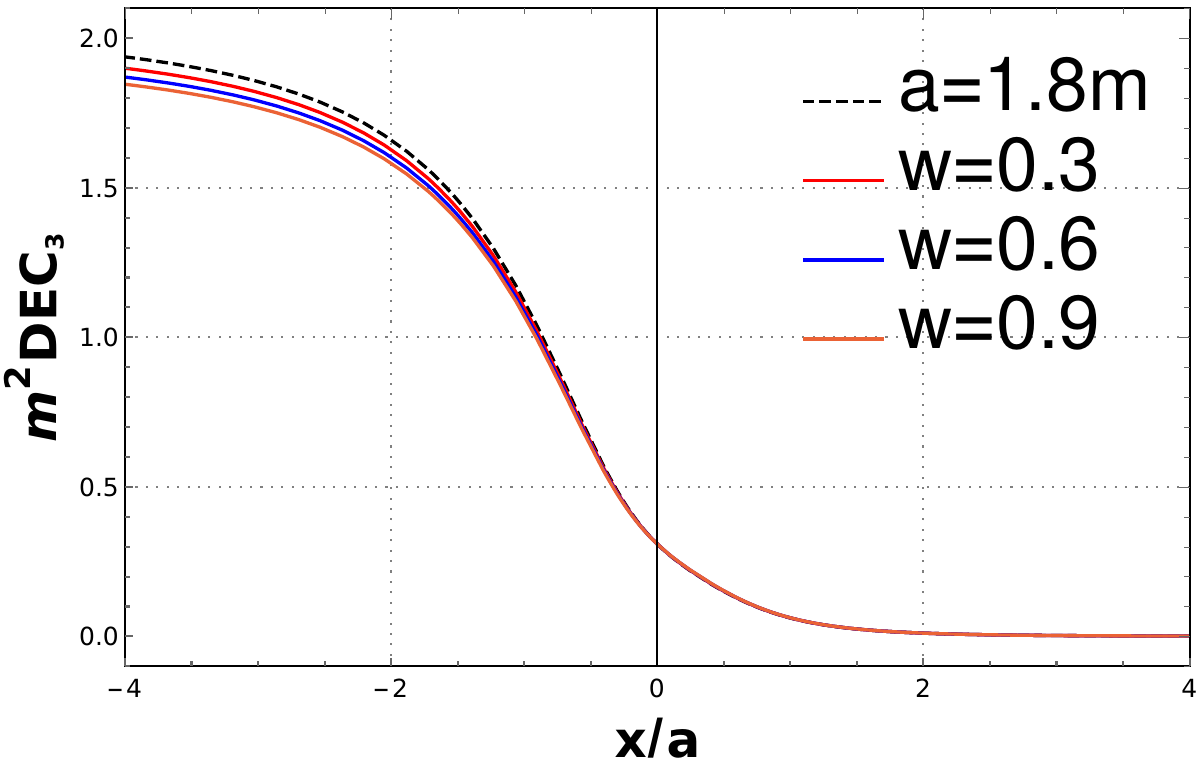}}}} 
\mbox{
	\subfigure[]{\label{DEC2FMOD1}
	{\includegraphics[width=0.65\linewidth]{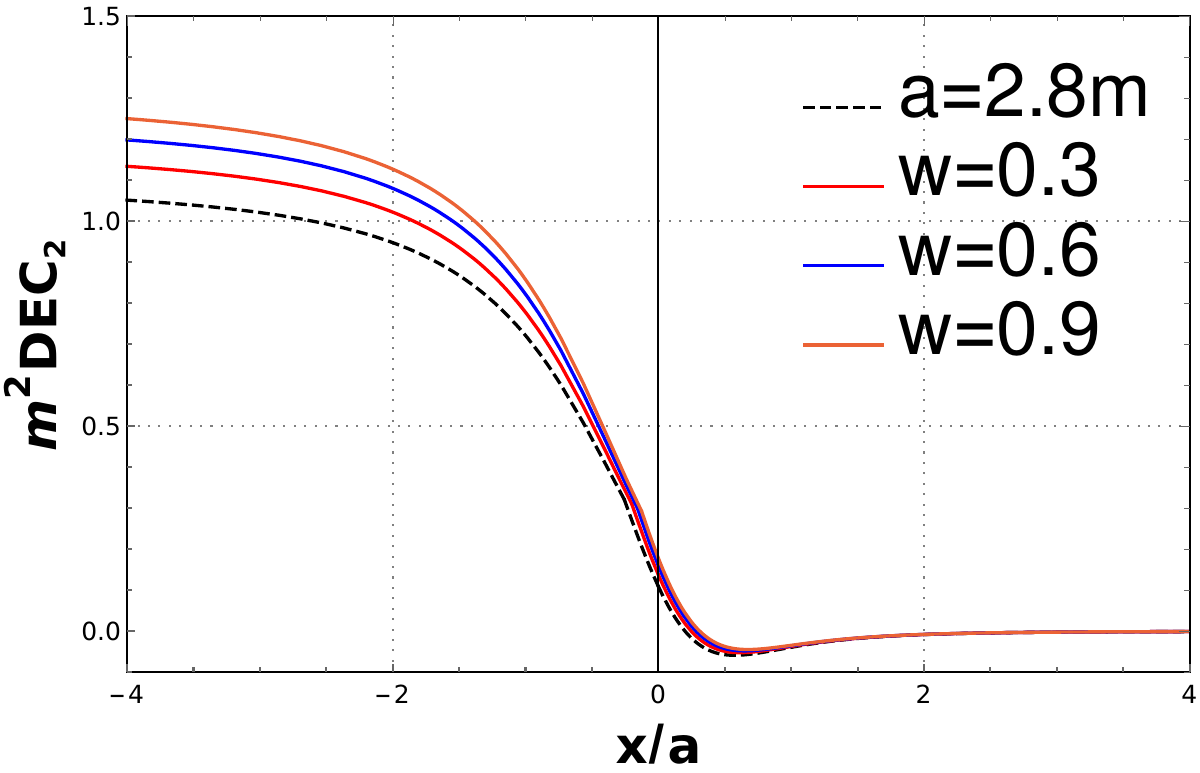}}} }
\caption{Complementary analysis of the energy conditions in which a throat radius was fixed inside and outside the event horizon, and then we varied the LSV parameter. The dotted black curve represents $W=0$. In (a), (c), and (e) we have, respectively, the inequalities
$SEC_3$, $DEC_3$, and $DEC_2$ to the cases where there is no horizon in $x>0$ while in (b) and (d) we have $SEC_3$ and $DEC_3$, respectively, to
the cases with horizons in $x>0$.}
\label{FIG6}
\end{figure}

In Fig. \ref{FIG6}, we have a complementary graphical representation for the energy conditions analyzed above considering the LSV factor, where the dotted black curve represents that the violation effect is turned off, $W=0$. We expect that the energy conditions could be alleviated by varying the LSV parameter so that some energy conditions would not be violated. However, this does not occur, thus maintaining the entire discussion already carried out in Refs. \cite{CDJM1,CDJM2}.

\subsection{Analysis of complementary energy conditions second model}\label{sec5b}

\begin{figure}[htb!]
\centering  
	\mbox{
	\subfigure[W=0]{\label{SEC3CMOD2}
	{\includegraphics[width=0.3\linewidth]{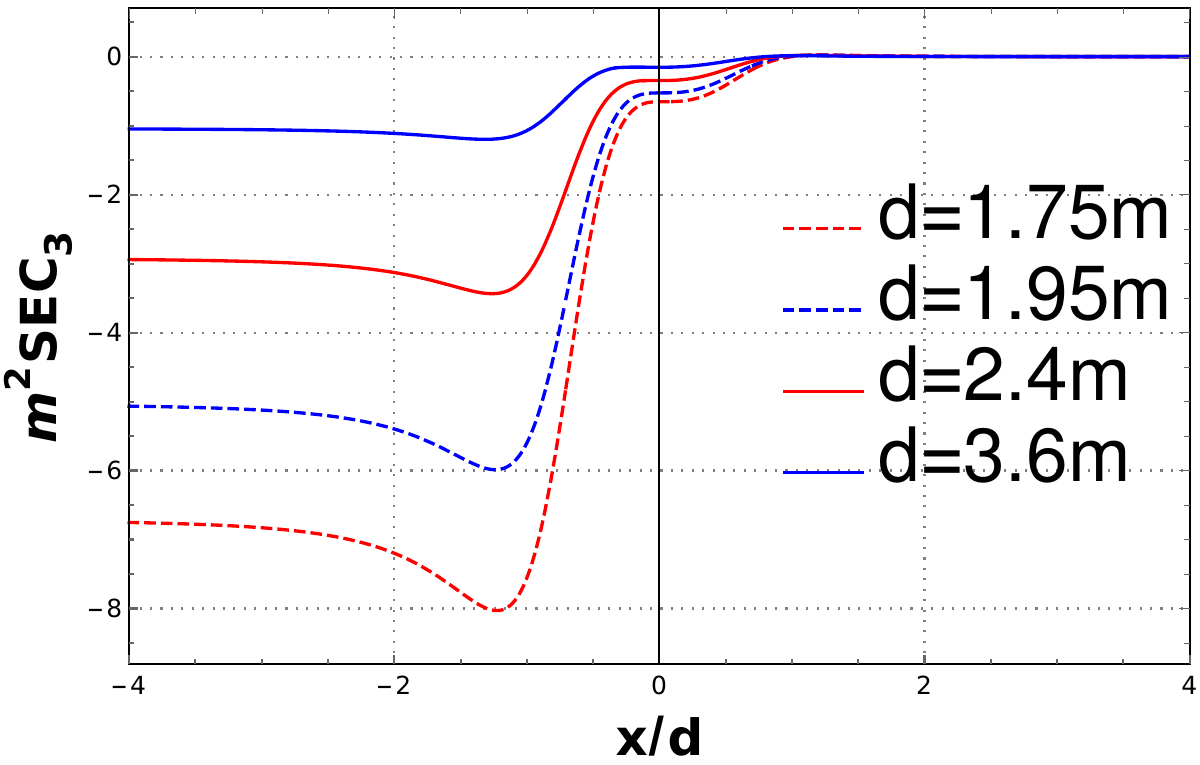}}}\qquad
	\subfigure[]{\label{SEC3DMOD2}
	{\includegraphics[width=0.3\linewidth]{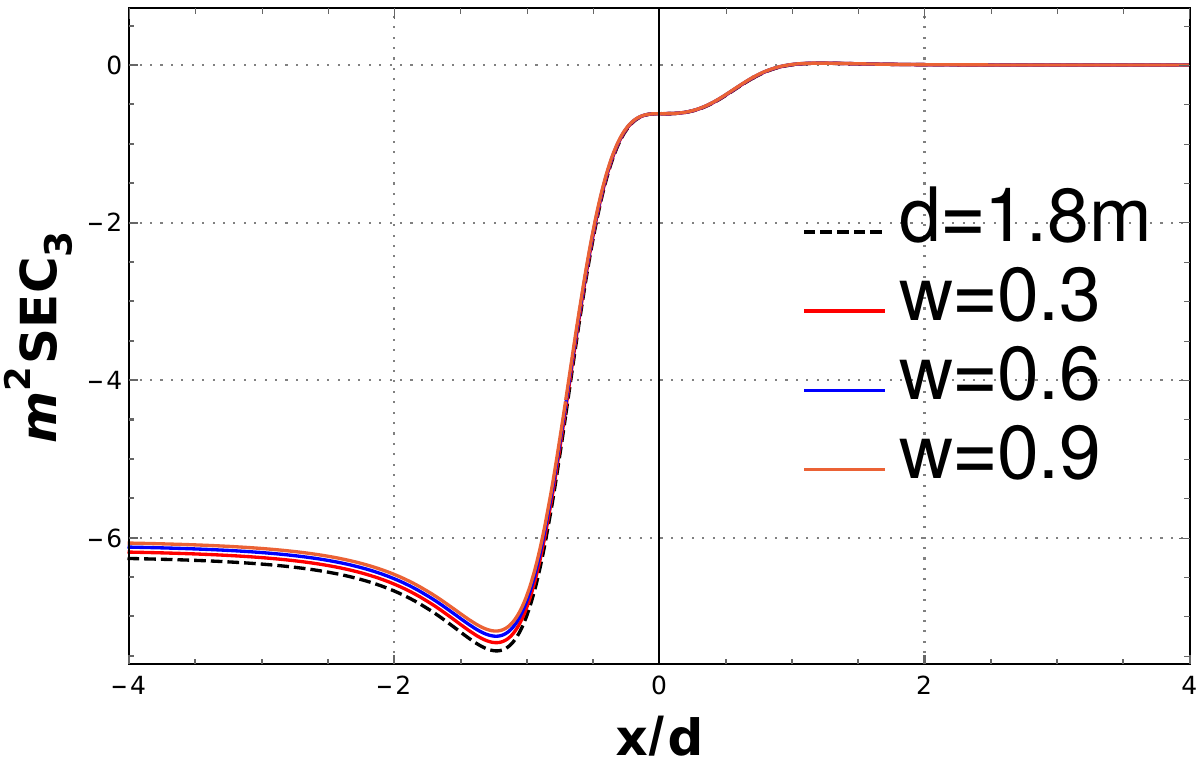}}}\qquad
    \subfigure[]{\label{SEC3FMOD2}
	{\includegraphics[width=0.3\linewidth]{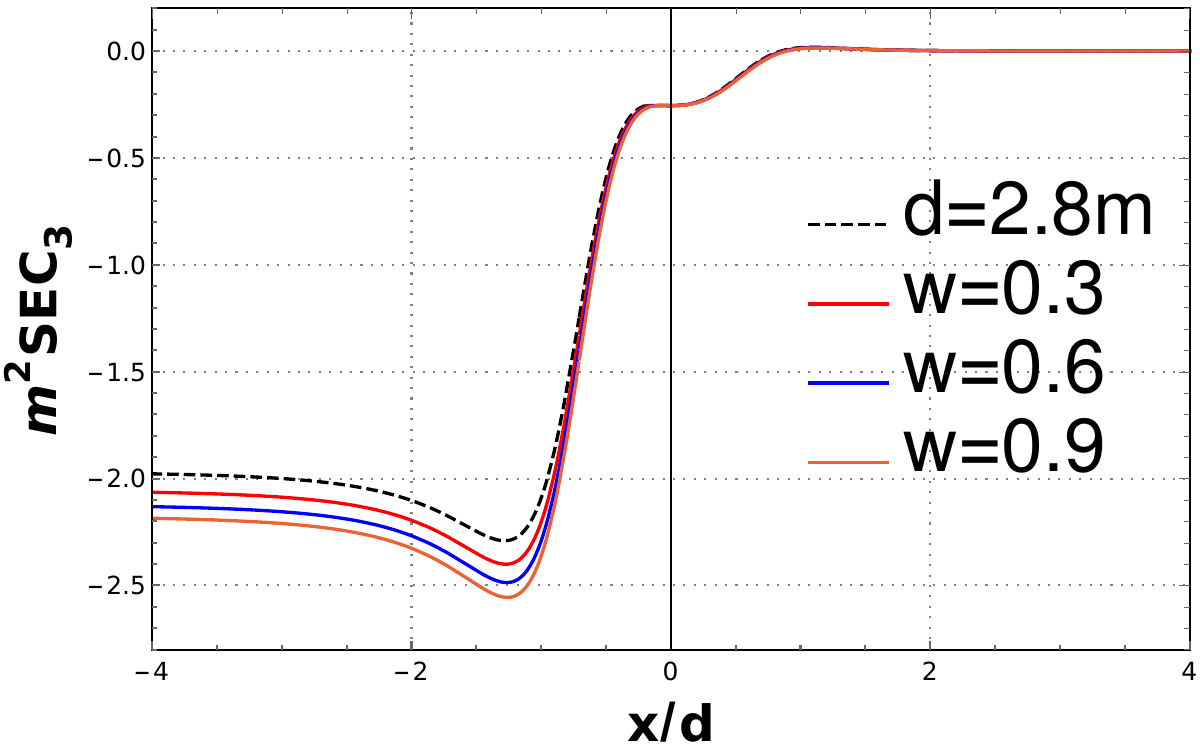}}}}
    \mbox{
	\subfigure[W=0]{\label{DEC3CMOD2}
	{\includegraphics[width=0.3\linewidth]{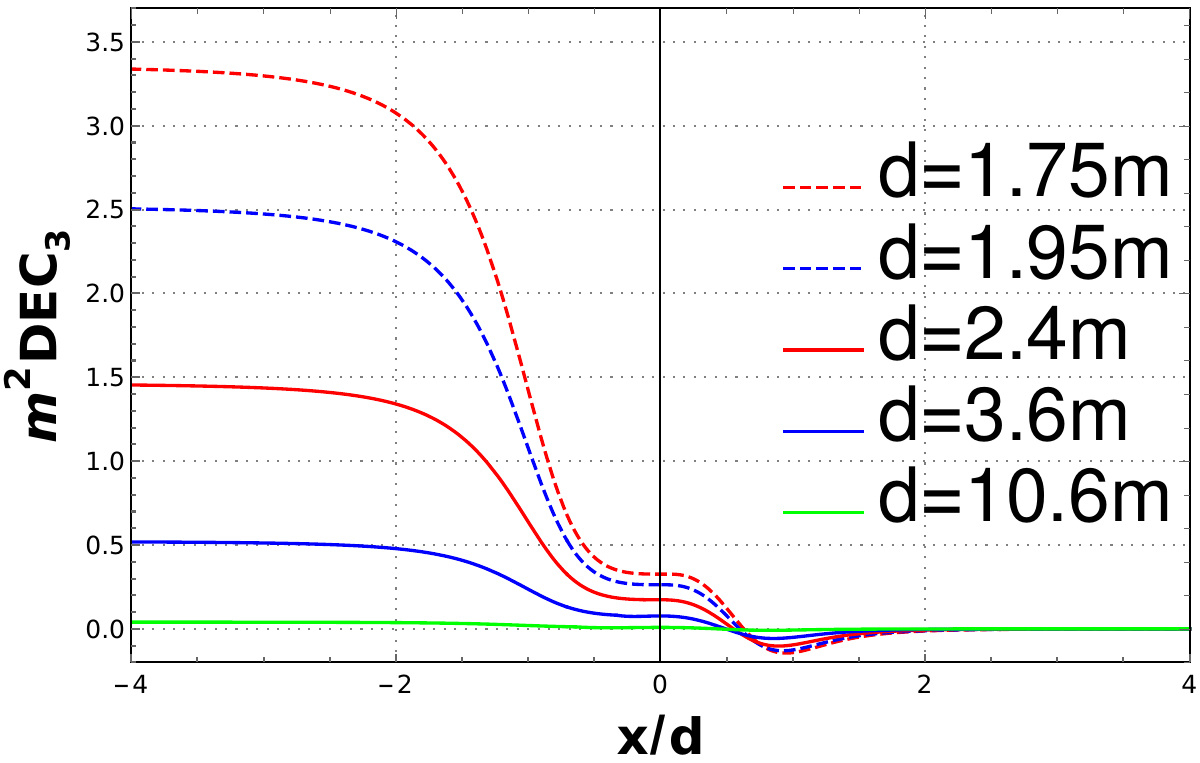}}}\qquad
	\subfigure[]{\label{DEC3DMOD2}
	{\includegraphics[width=0.3\linewidth]{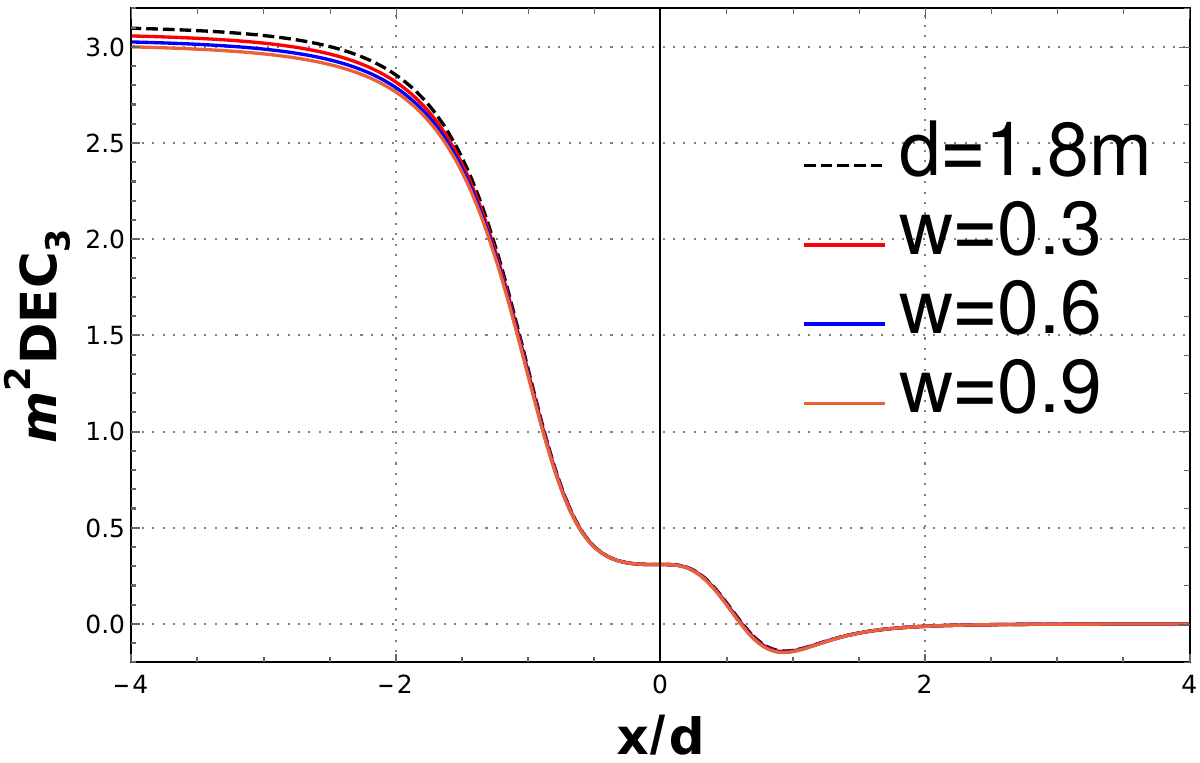}}}\qquad
    \subfigure[]{\label{DEC3FMOD2}
	{\includegraphics[width=0.3\linewidth]{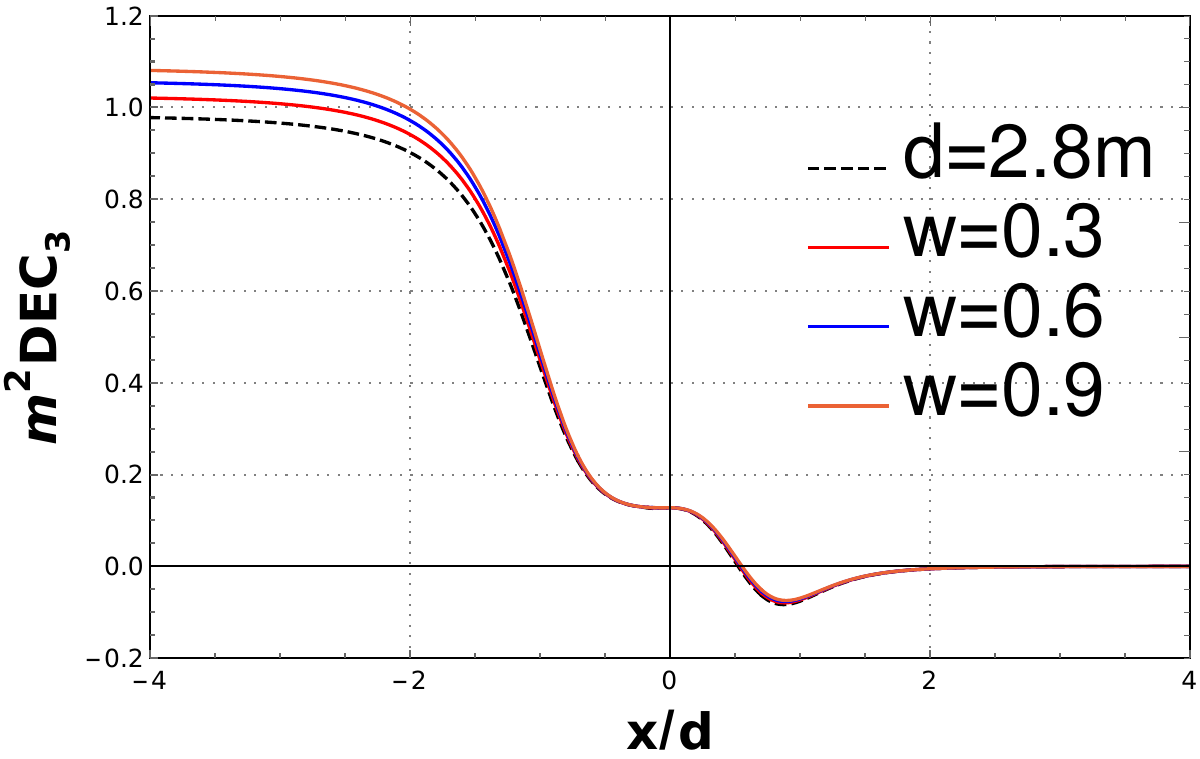}}}}
    \mbox{
	\subfigure[]{\label{DEC2FCMOD2}
	{\includegraphics[width=0.6\linewidth]{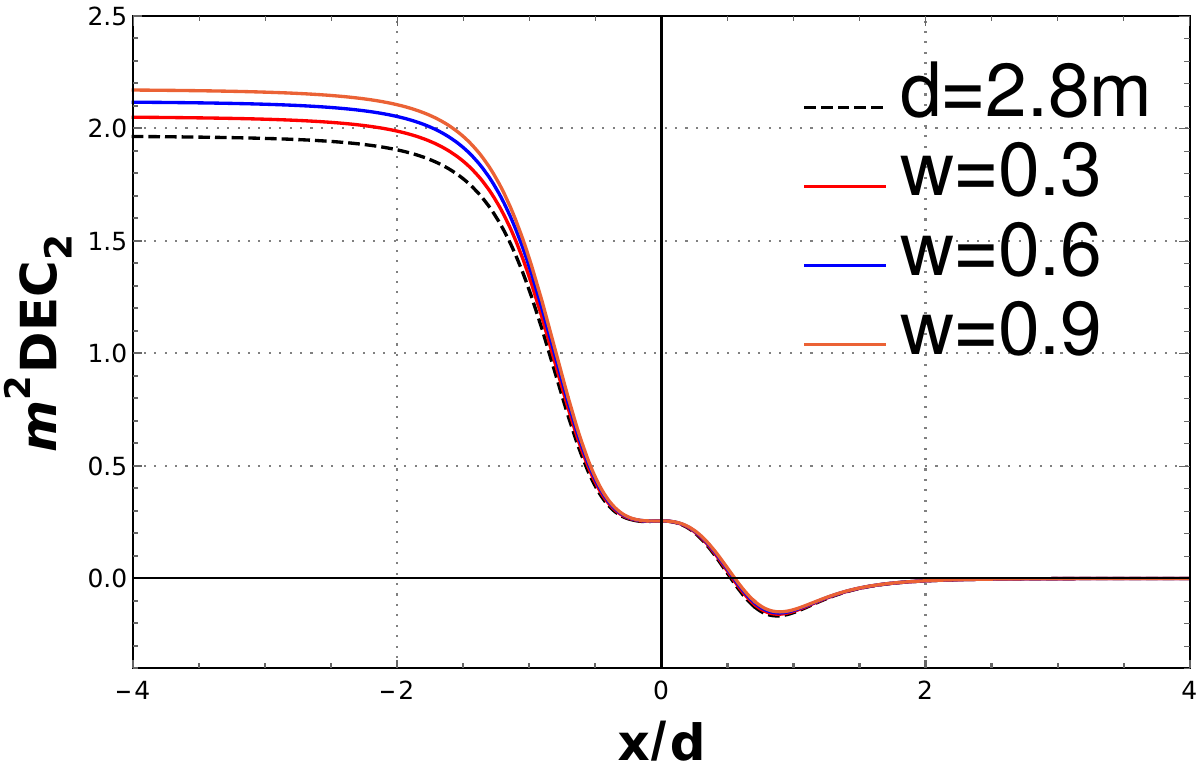}}}}
\caption{Complementary analysis of the energy conditions for the new model \ref{sec4}, with and without LSV effects. The dotted black curve represents $W=0$. In (b) and (e) we have, respectively, $SEC_3$ and $DEC_3$
to $d=1.8$. In (c), (f), and (g) we have, respectively, $SEC_3$, $DEC_3$, and $DEC_2$ to $d=2.8$.}
\label{FIG7}
\end{figure}

In this subsection, we will perform a complementary analysis of the energy conditions for which it was impossible to draw conclusions based solely on their expressions regarding the new black-bounce model \ref{sec4}. We can start by observing Fig. \ref{FIG7}, which contains all the information necessary to finalize conclusions regarding the violation of the energy conditions. In Fig. \ref{DEC2FCMOD2}, we have the dominant energy condition $DEC^\phi_2$ analyzed for throat radii outside the event horizon. Then, we vary the LSV parameter, which does not produce significant modifications. This complement to the secondary null energy condition $NEC^\phi_2$  Eq. (\ref{eq40})leads to the idea that both are violated, inside and outside the horizon.

In Fig. \ref{SEC3CMOD2}, we have that the strong energy condition, $SEC^\phi_3$, is violated for values of throat radii inside and outside the event horizon. In Figs. \ref{SEC3DMOD2} and \ref{SEC3FMOD2}, we fix a radius of the throat inside and outside the horizon and then vary the LSV parameter; however, no significant change can be noticed. Finally, in Fig. \ref{DEC3CMOD2}, we see that the dominant energy condition is not violated inside the horizon and is violated for radii outside. In Figs. \ref{DEC3DMOD2} and \ref{DEC3FMOD2}, we fix a radius of the throat inside and outside the horizon and then we vary the violation parameter, which does not produce significant changes in the energy conditions.

\section{Conclusion}\label{sec6}

In this work, we started from a spherically symmetric and static solution in a metric affine bumblebee gravity \cite{LV1,LV2}. This model can be initially used to measure the effects of LSV in a Schwarzschild spacetime, for example, and then verify how its main properties are modified by the violation parameter. Thus, we applied the same spacetime regularization procedure proposed by Simpson-Visser [19] to regularize this new metric tensor and then investigated possible black-bounce solutions that would be modified by LSV effects, in a model that had only the $k$-essence scalar field as a source.

We primarily analyzed the effects of LSV in the black-bounce model in a $k$-essence theory already investigated in \cite{CDJM1} for the configuration $n=1/3$, and then we analyzed its implications both for the metric function itself and for the other quantities derived directly from them (see Sec. \ref{sec3}). The function $A(x)$ tended to $\Pi$ as x $\to \infty$. Thus, we observed how the violation of Lorentz symmetry manifested at infinity. Finally, we concluded that new energy conditions now modified by the effects of LSV were not alleviated.

We presented a new black-bounce solution in a $k$-essence theory (see Sec. \ref{sec4}), but then considered a new area function $\Sigma^2_1$ and looked for the corresponding metric function. We obtained a general solution Eq. (\ref{eq24}) that had the behavior of being asymptotically de Sitter for $x\to\infty$ and anti-de Sitter for $x\to{-\infty}$. This area function was initially proposed and explored in Ref. \cite{MM1}. For our purposes, we applied two constraints on the general solution, requiring it to be asymptotically flat for $x\to\infty$ and recovering the Simpson-Visser limit at the origin $A(x\to{0})=1-\frac{2m}{d}$.

We constructed and analyzed the scalar and potential fields for the new model and investigated the possibility that LSV effects could relax the energy conditions to the point where there was no violation, but this did not occur.

\begin{acknowledgments}
We thank CNPq, CAPES, FUNCAP, and FAPES for financial support. 
\end{acknowledgments}

\nocite{*}
		
\end{document}